\documentclass[a4paper,fleqn,usenatbib,useAMS]{mnras}

\usepackage[utf8]{inputenc}
\usepackage{longtable}
\usepackage{graphicx}	
\usepackage{amsmath}	
\usepackage{amssymb}	
\usepackage{multicol}        
\usepackage{bm}		
\usepackage{pdflscape}	
\usepackage{xcolor}
\usepackage{soul}
\usepackage{footmisc}
\usepackage{float}

\usepackage{float}

\usepackage[T1]{fontenc}
\usepackage{ae,aecompl}

\title[Pre-main sequence variable stars]{Variable stars in the Sh 2-170 H\,{\sc ii} region }

\author[T. Sinha et al.]{Tirthendu Sinha$^{1,~6}$\thanks{E-mail: tirthendu@aries.res.in}, 
Saurabh Sharma$^{1}$\thanks{E-mail: saurabh@aries.res.in}, A. K. Pandey$^{1}$, R. K. Yadav$^{2}$,  K. Ogura$^{3}$,
\newauthor
N. Matsunaga$^{4}$, N. Kobayashi$^{5}$,  P. S. Bisht$^{6}$, R. Pandey$^{1}$ and A. Ghosh$^{1}$
\\
$^{1}$Aryabhatta Research Institute of Observational Sciences (ARIES), Manora Peak, Nainital, 263 002, India\\
$^{2}$National Astronomical Research Institute of Thailand, Chiang Mai 50200, Thailand\\
$^{3}$Kokugakuin University, Higashi, Shibuya-ku, Tokyo 150-8440, Japan\\
$^{4}$Department of Astronomy, School of Science, The University of Tokyo, 7-3-1 Hongo, Bunkyo-ku, Tokyo 113-0033, Japan\\
$^{5}$Institute of Astronomy, University of Tokyo, 2-21-1 Osawa, Mitaka, Tokyo 181-0015, Japan\\
$^{6}$Department of Physics, SSJ Campus, Kumaun University, Nainital 263 002, India\\
}

\date{ }

\pubyear{2019}

\begin{document}
\label{firstpage}
\pagerange{\pageref{firstpage}--\pageref{lastpage}}
\maketitle

\begin{abstract}
We present multi-epoch deep ($\sim$20 mag) $I_{c}$~band photometric monitoring of the Sh 2-170 star-forming region to understand the variability properties of pre-main-sequence (PMS) stars. We report identification of 47 periodic and 24 non-periodic variable stars with periods and amplitudes ranging from $\sim$4 hrs to 18 days and from $\sim$0.1 to 2.0 mag, respectively. We have further classified 49 variables as PMS stars (17  Class\,{\sc ii} and 32  Class\,{\sc iii}) and 17 as main-sequence (MS)/field star variables. A larger fraction of  MS/field variables (88\%) show periodic variability as compared to the PMS variables (59\%). The ages and masses of the PMS variable stars are found to be comparable with those of T-Tauri stars.  Their variability amplitudes show an increasing trend with the near-IR/mid-IR excess.  The period distribution of the PMS variables shows two peaks, one near $\sim$1.5 days and the other near $\sim$4.5 days.  It is found that the younger stars with thicker discs and envelopes seem to rotate slower than their older counterparts.  These properties of the PMS variables support the disc-locking mechanism.  Both the period and amplitude of PMS stars show decrease with increasing mass probably due to the effective dispersal of circumstellar discs in massive stars.  Our results favour the notion that cool spots on weak line T-Tauri stars are responsible for most  of their variations, while hot spots on classical T-Tauri stars resulting from variable mass accretion from an inner disc contribute to their larger amplitudes and irregular behaviours.
\end{abstract}





\section{Introduction}

 The evolution of pre-main-sequence (PMS) stars involves a set of complex physical processes such as evolution of circumstellar 
discs, accretion processes, bipolar jets, rotation properties etc. Because of these processes, the PMS stars  
show a wide range of luminosity variability in almost  all wavelength ranges from X-ray to infrared (IR) 
and their variability time scales range from a few minutes to years 
\citep{1989A&ARv...1..291A,2000AJ....120..349H}. 
Several mechanisms are known that 
induce photometric variability in PMS stars, e.g.,  irregular distribution of cool spots on stellar photospheres, variable hot spots,
 obscuration by dust, instability in discs, change in accretion rates, etc, \citep[][and references 
therein]{1994AJ....108.1906H}. The evolution of circumstellar discs and the accretion rates may play a prominent role in the 
non-periodic variability \citep{2002astro.ph.10520H,2006PASP..118.1390P,2010PASP..122..753P}, 
whereas periodic/quasi-periodic variability may be caused by rotation of stars having cool and hot spots on the photosphere. 
Weak-line T-Tauri stars (WTTSs) usually show periodic variability due to 
spot modulation of cool spots on their surface, whereas classical T-Tauri stars (CTTSs) 
show non-periodic variability due to the irregular accretion processes  from their thick disc. 
The interaction of CTTS itself with the thick circumstellar disc is also complex 
as both the accretion rate and the distribution 
of accretion zones (potential hot spots) over the stellar surfaces are not uniform 
\citep{2007prpl.conf..297H}. Intermediate mass counterparts of CTTSs are Herbig Ae/Be stars (spectral type A/B with 
emission lines). The presence of circumstellar patchy dust clouds causes photometric variation in most Herbig Ae/Be stars 
\citep{1998A&A...330..145V}. Apart from PMS variables, many MS stars also show
variability, e.g., $\beta$ Cep type stars, slowly pulsating B-type stars (SPB) and $\delta$ Scuti stype stars,
due to their pulsating behaviour \citep{2013A&A...554A.108M,2014MNRAS.442..273L}.

The period of spotted variables is the direct indicator of the rotational speed and hence they are related to the angular momentum. 
As the PMS stars contract towards the MS, they should increase their rotation speed up to  the break-up velocity due 
to the conservation of angular momentum. But numerous studies on PMS stars have revealed that they typically rotate at much 
smaller fraction of the  break-up velocity in spite of their significant contraction.  To explain the effective 
removal of the angular  momentum from PMS stars during the first $\sim$10 Myr of their evolution, several mechanisms have 
been proposed including magnetic star-disc interaction 
\citep[disc-locking,][]{1991ApJ...370L..39K,1994ApJ...429..797S,1995RMxAC...1..293N,1995ApJ...447..813O}, scaled-up solar-type 
magnetized winds driven by accretion 
\citep[e.g.,][]{2004ApJ...607L..43M, 2005ApJ...632L.135M, 2005MNRAS.356..167M, 2008ApJ...678.1109M, 2008ApJ...681..391M}, 
and scaled-up solar-type coronal mass ejection \citep[e.g.,][]{2009AIPC.1094..337A,2010ApJ...717...93A,2011SoPh..268..195A}. 
These mechanisms are still under debate and it is yet not clear to what extent each of them contributes
to the angular  momentum loss of PMS stars. Among them, the disc-locking mechanism has probably been verified from 
the bimodal distribution of variability period in young stars as the disc-locked slow 
rotators  can explain the separate period distribution
from the usual ones  \citep{1991ApJ...370L..39K,1994ApJ...429..797S}. This bimodality has been found in many young clusters  
with few exceptions \citep{2002A&A...396..513H,2004AJ....127.2228M}. The disc-locking mechanism can also be verified by checking 
the correlation 
of rotation periods with disc indicators such as: $\Delta (I_{c}-K_{s})$, $\Delta (H-K_{s})$, $[[3.6]-[4.5]]$, $H\alpha$ emission etc. 
\citep[][]{2002A&A...396..513H,2010A&A...515A..13R}.
Correlation of rotation period of PMS stars with different stellar properties (age, mass, accretion rate etc.)
can also be used to understand their disc evolution. \citet{2012MNRAS.427.1449L,2016MNRAS.456.2505L} have found a decrease 
of rotation period of PMS variables with the increase in their masses, 
but no conclusive result was found for age-period correlation. 
\citet{2018MNRAS.476.2813D} have found a correlation between variability amplitude with IR excess in PMS stars, 
indicating that in the disc bearing stars, the accretion phenomena plays a significant role in their energy output.
Unfortunately, we are still not able to construct a complete paradigm for disc evolution and a commonly adopted
scenario for early stellar evolution that satisfies all the available
observational constraints. This is why the interest to this subject does not fade out. 
Therefore, we have carried out long-term monitoring of the Galactic star-forming
region Sh 2-170 to study the variability properties of PMS stars and to understand their evolution.

\begin{figure*}
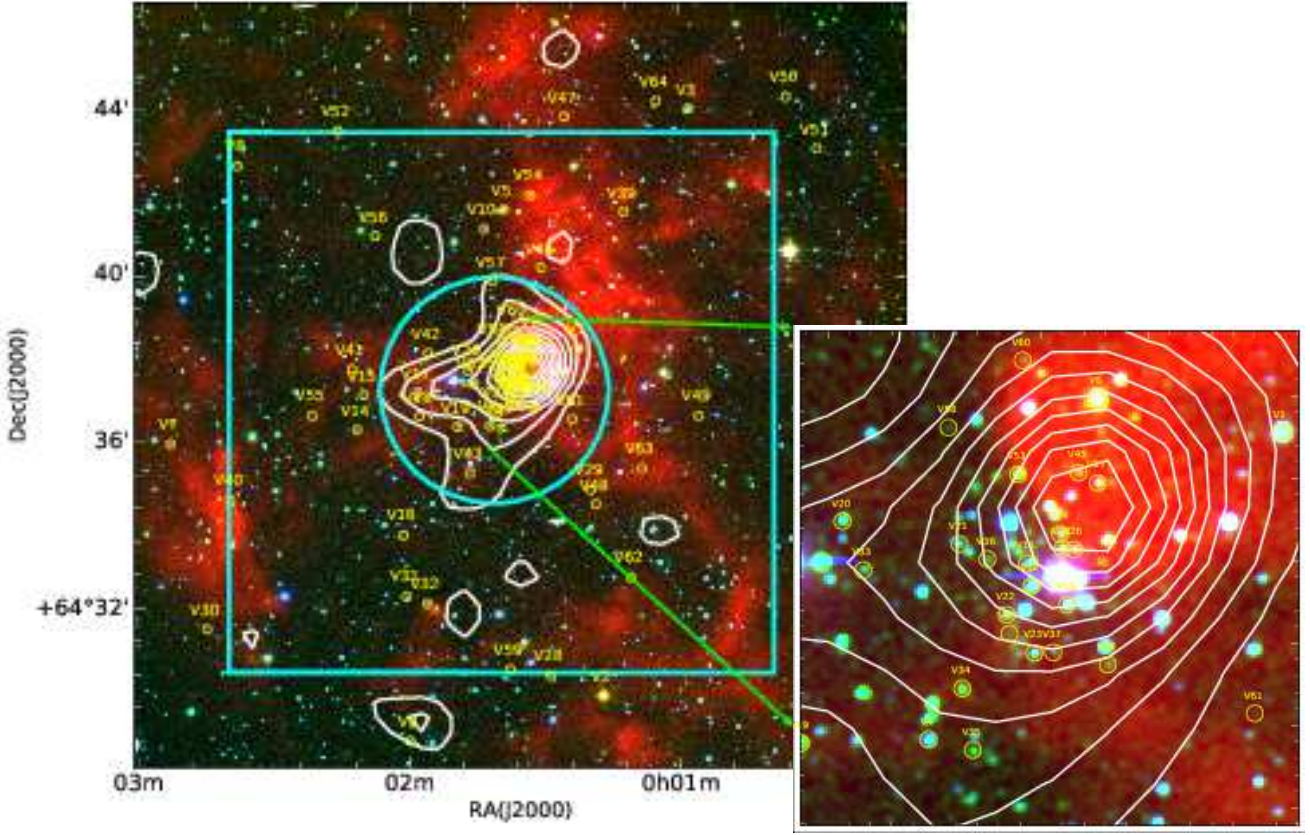

\centering
\hbox{
\hbox{
\includegraphics[width=0.68\textwidth]{figures/image.lreps}
}
\hbox{
\hspace{-2cm}
\includegraphics[width=0.38\textwidth]{figures/image-small.lreps}
}}
\caption{ Colour-composite image obtained by using the 0.8 $\mu$m $I_{c}$ (blue, present observations), 
2.17 $\mu$m $K_{s}$ (green, 2MASS), and  12 $\mu$m (red, WISE) 
images of the $\sim18^\prime.4\times 18^\prime.4$ FOV around the Sh 2-170 H\,{\sc ii} region.
The region observed by \citet{2012NewA...17..160B} is shown by a cyan box. The white contours represent the 
$K_{s}$ band stellar surface density distribution. The outermost contour represents the density 1$\sigma$ above the mean density
(i.e., 31$\pm$9 stars/arcmin$^2$)
and the step size is equal to 1$\sigma$ (9 stars/arcmin$^2$).
The cyan circle (radius = $2'.72$) indicates the extent of the cluster Stock 18. 
Identified variable stars are encircled  along with their identification numbers.
The inset is a zoomed-in image of the central region.
}
 \label{fig:image}
\end{figure*}

Sh 2-170 is an  H\,{\sc ii} region located at $\alpha_{2000}$ = 00$^{h}$01$^{m}$37$^{s}$, $\delta_{2000}$ = +64$^\circ$37$^{m}$30$^{s} 
($~$l$ = 117$^\circ$.62, $b$ = +2$^\circ$.27) in the Cassiopeia constellation with a diameter of $18'$ 
\citep{2004A&A...425..553R}. This H\,{\sc ii} region is excited by an 
O9V star, BD+63 2093p, which is a member of the star cluster Stock 18, situated at the centre of this
nebula  \citep{2007A&A...470..161R}.
Using optical two-colour diagrams (TCDs) and colour-magnitude diagrams (CMDs), \citet{2012NewA...17..160B} 
have estimated the age and distance of Stock 18  as 6 Myr and 2800 $\pm$ 200 pc, respectively.
They have also found that this region harbors a number of PMS stars with a span in their ages as well as in  masses, 
making it an ideal site to study their variability properties. 
We have done a long-term monitoring of this region using our deep and wide field optical observations taken with 
the 1-meter class telescopes located in India, China and Thailand.
This has been used, along with the recently available proper motion (PM) data from the {\it $Gaia$} second data release 
\citep{2018A&A...616A...1G,2016A&A...595A...1G} (DR2)
and near-IR (NIR)/mid-IR (MIR) photometric data from  2MASS, $Spitzer$ and WISE, 
 to make a detailed analysis of the light curves (LCs) and stellar properties of the
variable stars identified in this star-forming region.
In this paper, Section \ref{sect:obs} describes the observations and data reduction.
The stellar density distribution, membership probability,
age/distance of this region along with the identification of variables
and  derivation of their physical parameters
are presented in Section \ref{sect:Result}.
The characteristics  of the variables and the correlation of their period and amplitude of variability
with their physical parameters are discussed in Section \ref{sect:discussion}.
We conclude our studies in Section \ref{sect:conclusion}.

\begin{figure*}
\centering
\includegraphics[width=0.52\textwidth, angle=0]{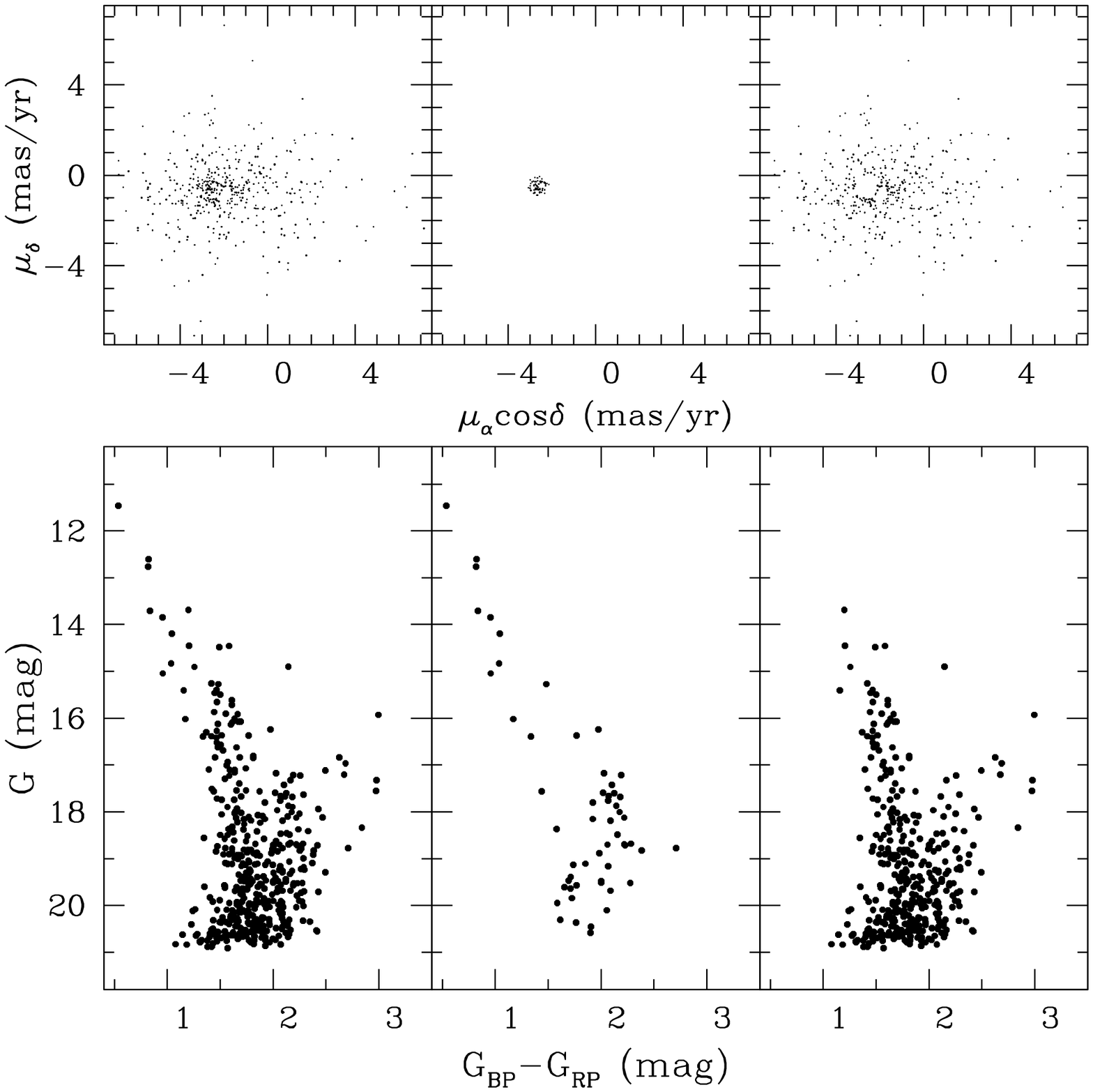}
\hspace{0.1 cm}
\includegraphics[width=0.39\textwidth, angle= 0]{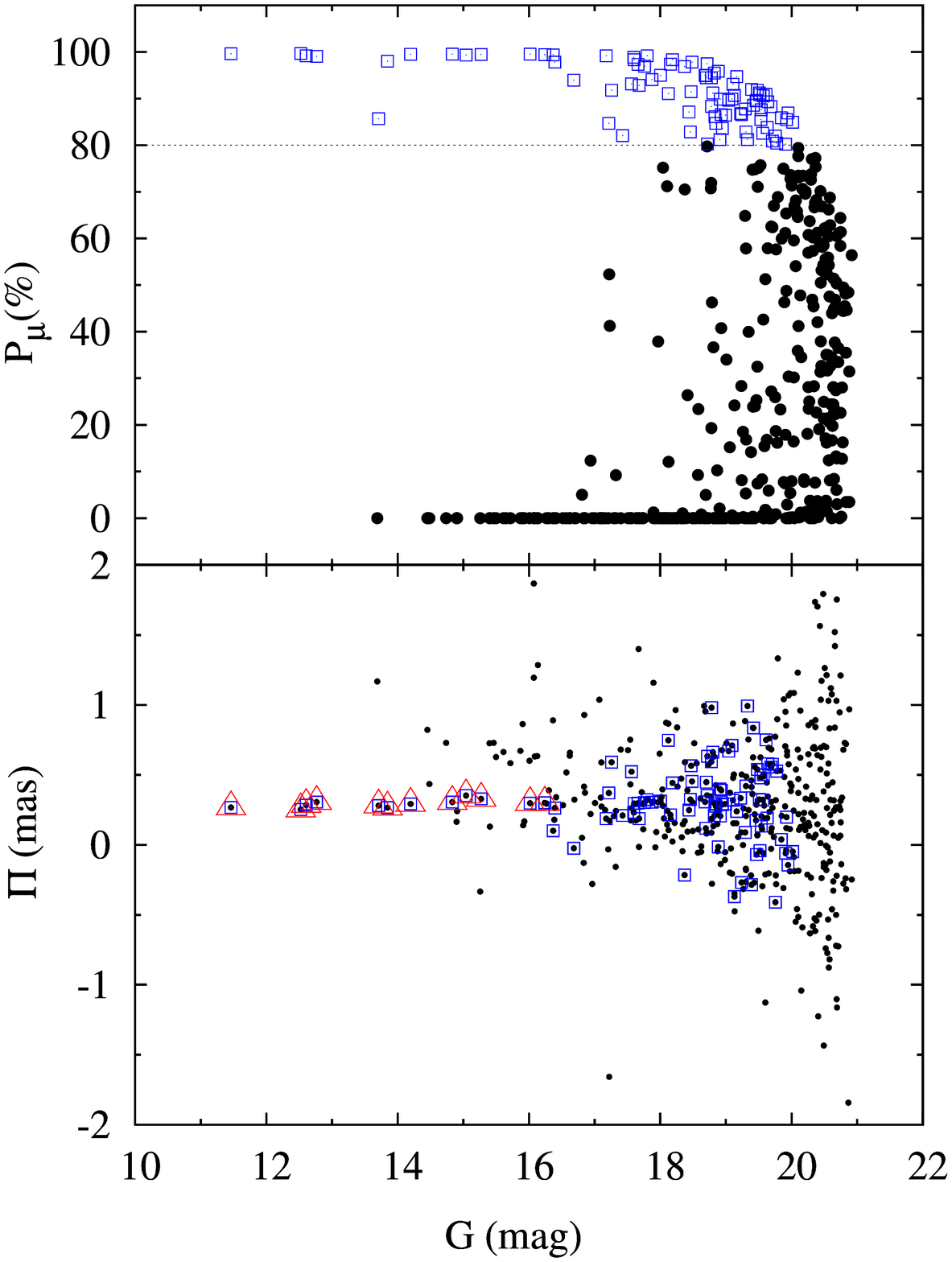}
\caption{ Left panel: Proper motion vector-point diagrams (VPDs; top sub-panels) and
{\it $Gaia$} DR2 $G$/$(G_{BP} - G_{RP})$ CMDs (bottom sub-panels) for the stars located in the central
 cluster Stock 18 region (r < 2$^\prime$.72).
The left sub-panels show all the stars, while the middle and right sub-panels show
probable cluster members and field stars.
Right panel: Membership probability (P$_\mu$) and parallax ($\Pi$) as a function of $G$ magnitude for
the stars located in the central cluster (black dots).
The probable member stars (P$_\mu\geq$ 80\%) are shown by blue squares in both sub-panels.
Red triangles indicate  probable member stars with parallax errors better than 0.05 mas.}
 \label{fig: VPD}
\end{figure*}

\begin{tiny}
\begin{table}
 \caption{Log of observations.}
 \label{tab:anysymbols}
 \begin{tabular}{cccc}
  \hline
Date & Telescope & No. of frames $\times$        & Filter \\
     &           &      exposure time (sec)     &        \\
  \hline 
30.09.2016    & DFOT  &   085$\times$180       &$I_{c}$   \\ 
              &       &   003$\times$180       &$V$   \\ 
21.10.2016    & DFOT  &   135$\times$180       &$I_{c}$   \\ 
              &       &   003$\times$180       &$V$   \\ 
11.11.2016    & DFOT  &   085$\times$180       &$I_{c}$   \\ 
              &       &   003$\times$180       &$V$   \\ 
26.11.2016    & DFOT  &   020$\times$180       &$I_{c}$   \\ 
              &       &   003$\times$180       &$V$   \\ 
17.12.2016    & DFOT  &   001$\times$180       &$I_{c}$   \\ 
              &       &   001$\times$180       &$V$   \\ 
23.12.2016    & TRT-GAO & 016$\times$240        &$I_{c}$   \\ 
24.12.2016    & TRT-GAO & 026$\times$240        &$I_{c}$   \\ 
24.12.2016    & 0.5-m TNO & 023$\times$300        &$I_{c}$   \\ 
26.12.2016    & DFOT  &   001$\times$180       &$I_{c}$   \\ 
14.10.2017    & DFOT  &   044$\times$180       &$I_{c}$   \\
15.10.2017    & DFOT  &   120$\times$180       &$I_{c}$   \\
17.10.2017    & DFOT  &   089$\times$180       &$I_{c}$   \\
18.10.2017    & DFOT  &   020$\times$180       &$I_{c}$   \\
25.10.2017    & DFOT  &   025$\times$180       &$I_{c}$   \\
26.10.2017    & DFOT  &   020$\times$180       &$I_{c}$   \\
21.11.2017    & DFOT  &   040$\times$180       &$I_{c}$   \\ 
24.11.2017    & DFOT  &   010$\times$180       &$I_{c}$   \\ \hline
\end{tabular}
\end{table}
\end{tiny}

\section{Observation and data reduction}
\label{sect:obs}

\subsection{Optical photometric data}

Optical photometric observations of the Sh 2-170 region were taken in the $I_{c}$ band on 17 nights 
and in the $V$ band on 5 nights starting from 30th September 2016 to 24th November 2017 
with the 1.3m Devasthal Fast Optical Telescope (DFOT, India), 0.7m Thai Robotic Telescope (TRT-GAO, Gao Mei Gu Observatory, China) 
and 0.5m telescope of Thai National Observatory (TNO, Thailand).
All the telescopes have a 2048$ \times $2048 pixel square CCD for imaging. 
The 1.3m Devasthal telescope covers a field of view (FOV) of $18^{\prime}.4 \times 18^{\prime}.4$,  
whereas the 0.7m and 0.5m telescopes have a FOV of $\sim20^{\prime}.9\times20^{\prime}.9$
and $\sim23^{\prime}\times23^{\prime}$, respectively. 
In total, 776 and 13 frames were taken in $I_{c}$ and $V$ filters, respectively, with exposure times of 180s on the 
1.3m, 240s on the 0.7m and 300s on the 0.5m telescopes. 
Bias and flat frames were also taken in each night along with the target frames. 
The typical seeing size (estimated from the FWHM of stellar images) during the observations on 1.3m DFOT were about 
$1^{\prime \prime}.5$ - $2^{\prime \prime}$ 
and for the 0.5m and 0.7m telescopes, it was $3^{\prime \prime}$ - $4^{\prime \prime}$. 
Details of the  observations are given in Table \ref{tab:anysymbols}. In Fig. \ref{fig:image}, we have shown the colour-composite 
image of the observed region by using the 12 $\mu$m (WISE, red colour), 2.17 $\mu$m  ($K_{s}$, 2MASS, green colour)  and 
0.8 $\mu$m  ($I_{c}$, present observation, blue colour) images.

The basic image processing, such as bias subtraction, flat fielding, cosmic ray rejection, 
were done by using the tasks available within IRAF\footnote{IRAF
is distributed by the National Optical Astronomy Observatory, which
is operated by the Association of Universities for Research in Astronomy (AURA) under cooperative agreement with the National Science
Foundation.}. The instrumental magnitude was obtained by using the DAOPHOT \citep{1987PASP...99..191S} package. 
As the cluster region is very crowded, we carried out PSF photometry to get the magnitudes of the stars. 
We have used the DAOMATCH and DAOMASTER  routine of DAOPHOT-II \citep{1992ASPC...25..297S} 
to estimate the shifts in individual frames with respect to a reference frame and to 
get the magnitudes of stars detected in different frames. The CCD pixel coordinates of the stars have been converted to celestial 
coordinates (RA and Dec) for the J2000 epoch by using the \textit{GAIA software\footnote{http://star-www.dur.ac.uk/~pdraper/gaia/gaia.html}}.
The observations obtained with different telescopes were cross-matched by using their astrometry within a 1$^{\prime \prime}$
search radius.

We have calibrated the instrumental magnitudes into the standard system using the photometric data 
published by \citet{2012NewA...17..160B} 
for the $\sim13^{\prime}\times13^{\prime}$ FOV (shown with the cyan square box in Fig. \ref{fig:image})
of the cluster region.
The transformation equations used for photometric calibration are as given below:

\begin{equation}
(V-I_{c}) = (0.903 \pm 0.005)\times(v-i_{c}) + (0.019 \pm 0.009)
\end{equation}
\vspace{-0.5 cm}
\begin{equation}
(V-v) = (-0.039 \pm 0.004)\times(V-I_{c}) + (2.956 \pm 0.008)
\end{equation}

where $V$, $I_{c}$ and  $v$, $i_{c}$ are standard magnitudes and instrumental magnitudes, respectively. 

In addition to the data from the present observations, we have used the available photometric data from 
\citet{2012NewA...17..160B} for our analysis.  
The catalog by \citet{2012NewA...17..160B} provides data for 2261 stars in a FOV of $\sim 13^\prime\times13^\prime$.
We have detected extra 4103 stars in a total FOV of $\sim 18^\prime.4\times18^\prime.4$ in the Sh 2-170 region.

\subsection{Archival NIR/MIR photometric data}
\label{NIR/MIR data}

 Since NIR and MIR data are very useful  to study the spectral energy distribution (SED) 
and disc properties of young stellar objects (YSOs), we have used  NIR/MIR photometric data from available archives described 
below:\\

(i) NIR JHKs photometric data have been taken from  the 2MASS All-Sky Point Source Catalog 
\citep{2006AJ....131.1163S,2003yCat.2246....0C}.

(ii) Spitzer-IRAC observations at 3.6 and 4.5 $\mu$m have been taken from the GLIMPSE360 Catalog and Archive 
\citep{2004ApJS..154....1W}.

(iii) MIR data at  3.4, 4.6, 12 and  22 $\mu$m have been taken from  the Wide-field Infrared Survey Explorer (WISE) All-sky Survey 
Data release \citep{2010AJ....140.1868W}. 
  
The NIR/MIR data having photometric error $\leq 0.2$ mag and a matching radius of $1''$ were used to identify 
their optical counterparts.

\section{Results}
\label{sect:Result}
\subsection{Stellar distribution}

To study the stellar density distribution in the Sh 2-170 region, 
we have obtained a surface density map for a sample of stars taken from the 2MASS 
All-Sky Point Source Catalog (less affected by gas and dust distribution), 
covering $18^{\prime}.4 \times 18^{\prime}.4$ FOV around this region. 
We have generated a surface density map using the nearest neighbour (NN) method as 
described by \citet{2005ApJ...632..397G,2009ApJS..184...18G}. We took the radial distance necessary to 
encompass the $20^{th}$ nearest star and computed the local surface density in a grid size of 
11$^{\prime \prime}$, which was then smoothened to a grid size of $3\times3$ pixel$^2$. The density contours 
derived are plotted in Fig. \ref{fig:image} as white curves.
The lowest contour is 1$\sigma$ above the mean stellar density (i.e., 31$\pm$9 stars/arcmin$^2$) 
and the step size is equal to 1$\sigma$ (9 stars/arcmin$^2$). 
The stellar density enhancement in the centre region of Sh 2-170, which is the Stock 18 cluster, can
be easily seen from the contours. 
The core of the cluster is almost circular, whereas the outer contours are elongated.  
The density distribution shows a peak at $\alpha_{2000}$: $00^h01^m34^s.4$, $\delta_{J2000}$: $+64^\circ37^\prime48^{\prime \prime}$  
with a core radius (defined as the point where density becomes half of the peak density)  
of $\sim50^{\prime \prime}$. The extent of the cluster is shown with a circle having a  
radius of $\sim$2$^\prime$.72.  On the basis of the radial density profile using 2MASS data, 
\citet{2012NewA...17..160B} have reported  the core and cluster radius as 
18$^{\prime \prime}$ and 3$^\prime$.5, respectively.

\subsection{Membership}
\label{membership}

The new and precise parallax measurements up to a very faint magnitude limits ($G$ (330-1050 nm) $\approx$ 21 mag) by $Gaia$ DR2 
have opened a new dimension in the studies of membership determination in star clusters. 
The catalog can be queried on the $Gaia$ archive by using  {\textit ADQL at http://gea.esac.esa.int/archive/}.
$Gaia$ PM data have been used to determine the membership probability of stars 
located within the Stock 18 cluster region (radius = 2$^\prime$.72). 
The PMs $\mu_\alpha$cos($\delta$) and $\mu_\delta$  are plotted as a vector-point diagram (VPD) 
in the top sub-panels of Fig. \ref{fig: VPD}  (left panel). 
 The bottom sub-panels show the corresponding $G$/($G_{BP} - G_{RP}$) CMDs, where
$G_{BP}$ (330-680 nm) and $G_{RP}$ (630-1050 nm) covers the blue part and red part of the $G$~band.
The left sub-panels show all stars, while the middle and right sub-panels show
the probable cluster members and field stars, respectively. 
There seems to be an obvious clustering around $\mu_\alpha$cos($\delta$) = -2.62 mas yr$^{-1}$
and $\mu_\delta$ = -0.49 mas yr$^{-1}$. 
A circular area having a radius of 0.51 mas yr$^{-1}$ around the cluster centroid in the 
VPD seems to define the PMs of the cluster. The chosen radius is a compromise between the exclusion of 
cluster members with poor PMs and the inclusion of field stars sharing the cluster mean PM.

 Assuming a distance of 2.8 kpc for Stock 18 \citep[cf.][]{2012NewA...17..160B} and a radial velocity 
dispersion of 1 kms$^{-1}$ for open clusters \citep{1989AJ.....98..227G}, 
the expected dispersion, $\sigma_c$,  in the PMs of the Stock 18 members would be 0.075 mas yr$^{-1}$ . 
For the remaining stars in the region (i.e., probable field stars), 
we have calculated $\mu_{xf}$ = -1.12 mas yr$^{-1}$, $\mu_{yf}$ = -0.67 mas yr$^{-1}$ , 
$\sigma_{xf}$ = 6.22 mas yr$^{-1}$ and $\sigma_{yf}$ = 2.68 mas yr$^{-1}$ (where $\mu_{xf}$ and $\mu_{yf}$ are the field 
PM centre and  $\sigma_{xf}$ and $\sigma_{yf}$ are the field intrinsic PM dispersion).
These values are further used to construct the frequency distributions of the cluster stars ($\phi^\nu_c$ ) and field stars 
($\phi^\nu_f$ ). By using the procedure described by \citet{2013MNRAS.430.3350Y} the membership probability is estimated  
as per the  following equation,    

\begin{equation}
P_\mu(i) = {{n_c\times\phi^\nu_c(i)}\over{n_c\times\phi^\nu_c(i)+n_f\times\phi^\nu_f(i)}}
\end{equation}

 where $n_{c}$ (=0.19) and $n_{f}$ (=0.81) are the fractions of the cluster members and that of field stars, respectively. 
The membership probability of the sources within the cluster region (r < 2$^\prime$.72) 
is plotted as a function of $G$ magnitude in the top sub-panel of Fig. \ref{fig: VPD} (right panel). 
As can be seen a high membership probability (P$_\mu \geqslant$ 80\%) extends down to $G$ $\sim$ 20 mag.  
The bottom sub-panel of Fig. \ref{fig: VPD} (right panel) displays the parallax of  the 
same stars as a function of $G$ magnitude. Except few outliers, most of the stars with high membership 
probability (P$_\mu \geqslant$ 80\%) follow a tight distribution. 
We estimated the membership probability for 463 stars in the cluster region and
found 86 stars as cluster members (P$_\mu \geqslant$ 80\%).
The details of these members are given in Table \ref{PMT}.

\subsection{Distance and age}
\label{distace and age of cluster}

\citet{2018AJ....156...58B} have estimated distances of 1.33 billion stars using the 
data published in the $Gaia$ DR2 and those can be downloaded from $VizieR$\footnote{http://vizier.u-strasbg.fr/viz-bin/VizieR?-source=I/347\&-to=3}.
The individual distances of the 12 identified cluster members having parallax values with high accuracy 
(i.e. $error < 0.05$ mas, shown as red triangles in the bottom sub-panel of Fig. \ref{fig: VPD} (right panel)), 
have been obtained from \citet{2018AJ....156...58B}. The mean distance of  these members comes 
out to be $3.1\pm0.2$ kpc. This distance estimation is comparable, within the error, to that obtained by
\citet[][2.8$\pm$0.2 kpc]{2012NewA...17..160B} based on the $V/(B-V)$ CMD.

\begin{figure}
\centering
\includegraphics[width=0.9\columnwidth, angle= 0]{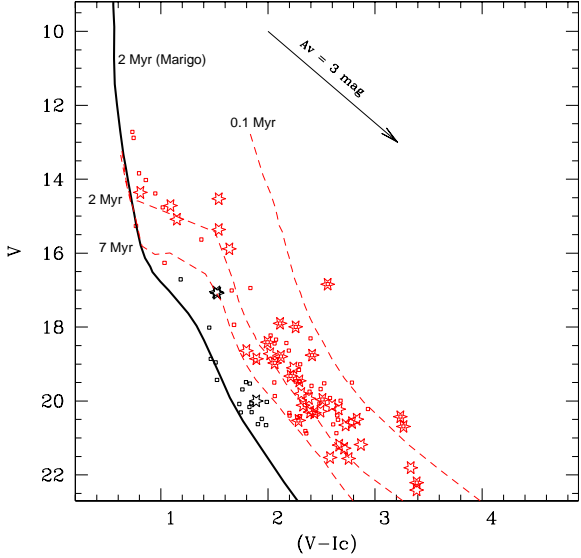}
\caption{ $V/(V-I_{c})$ CMD for the probable members (P$_\mu \geqslant$ 80\%) in the cluster region 
and the identified YSOs. The members are shown with open squares and YSOs with star symbols. 
The black thick curve shows 
the 2 Myr post-MS isochrone for the solar metallicity by \citet{2008A&A...482..883M} while the red dashed curves 
represent the PMS isochrones for 0.1, 2 and 7 Myr by \citet{2000A&A...358..593S}. The arrow indicates the reddening vector.
The symbols with red colour are for those stars associated with Sh 2-170 (see text for details).}
 \label{fig : MY_CMD}
\end{figure}

Fig. \ref{fig : MY_CMD} displays the $V/(V-I_{c})$ CMD of the member stars in the Stock 18 cluster 
(radius $<$ 2$^\prime$.72 and having P$_\mu$ $\geq 80\%$; 77 stars) shown by open square symbols. 
The post-MS isochrone for 2 Myr  for the solar metallicity  (black thick curve) by \citet{2008A&A...482..883M} 
along with the PMS isochrones of 0.1, 2, 7 Myr (red dashed curves) by \citet[][]{2000A&A...358..593S} are also shown. 
All the isochrones and evolutionary tracks are corrected for the distance of  3.1 kpc and for the reddening  
$E(V-I_{c})$ = 0.88 mag \citep[$E(B-V)$= 0.7 mag,][]{2012NewA...17..160B}.
The  $V/(V-I_{c})$ CMD indicates that a majority of the probable members are PMS stars. 
Few of the faint `members' located near the MS isochrones  at $V$ $\sim$ 20.5 mag are 
probably mis-identified stars due to large error in their parallax values near fainter  magnitude limits 
(cf. Fig. \ref{fig: VPD}). 
To investigate further about the age of the cluster, we have used NIR and MIR data from the 
2MASS, WISE and $Spitzer$ surveys (cf. Section \ref{NIR/MIR data}) to identify PMS stars showing excess 
IR emission, in the observed $18^{\prime}.4 \times 18^{\prime}.4$ area of the Sh 2-170 region.  We have used 
similar schemes as used in the recent literature \citep[]{2016AJ....151..126S,2014ApJ...791..131K,2009ApJS..184...18G}. 
On the basis of NIR/MIR excess emission we have identified 66 Class\,{\sc ii} YSOs in the region. 
A comparison with the optical data within 1$^{\prime\prime}$ matching radius yields optical counterparts for 
46 YSOs. The optical counterparts of Class\,{\sc ii} YSOs are also plotted as star symbols in Fig. \ref{fig : MY_CMD}. 
The locations of the YSOs in the CMD indicate that a majority of them are 
younger than 2 Myr with an upper age limit of 7 Myr, confirming the youth of this region. 
Thus, based on the distribution of the YSOs and most of the members in the optical CMD,
we define an upper age limit of 7 Myr for stars associated with Sh 2-170. These are shown with red symbols in 
Fig. \ref{fig : MY_CMD}.

\begin{figure}
\centering
\includegraphics[width=0.8\columnwidth, angle= 0]{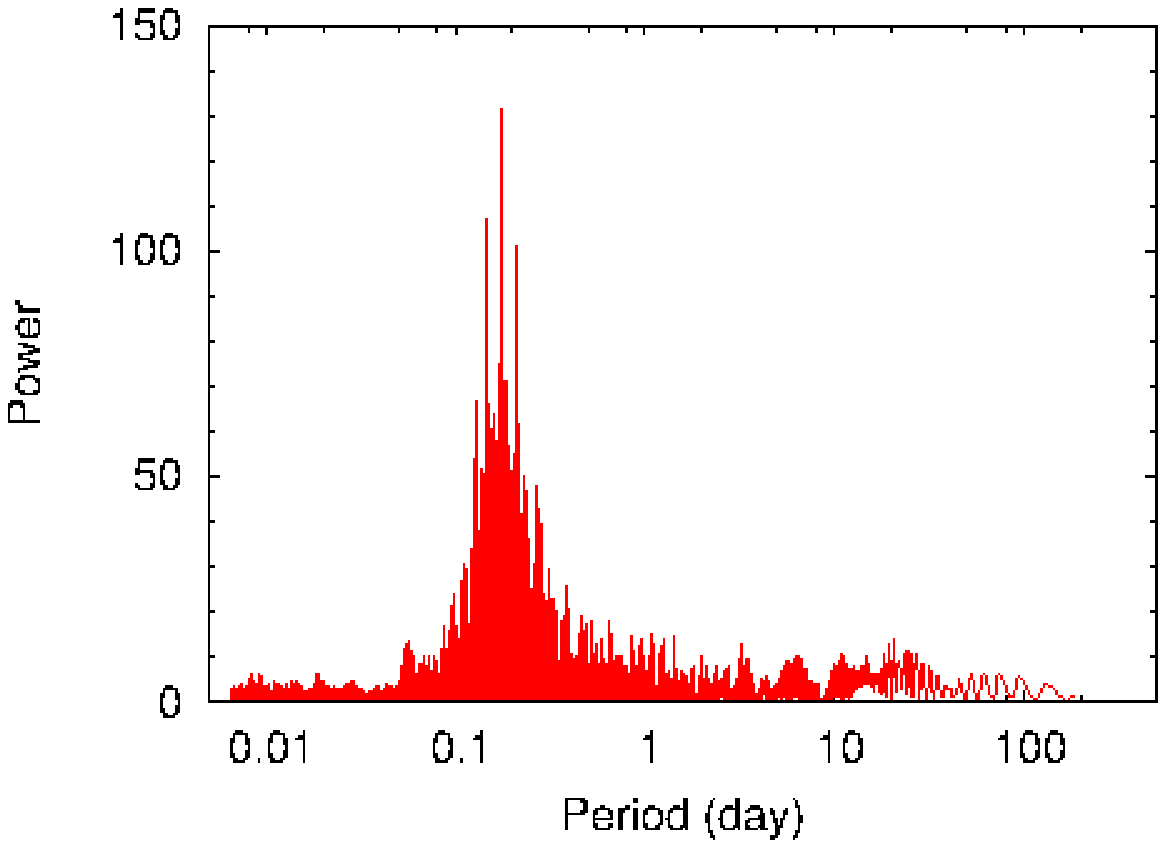}
\includegraphics[width=0.8\columnwidth, angle= 0]{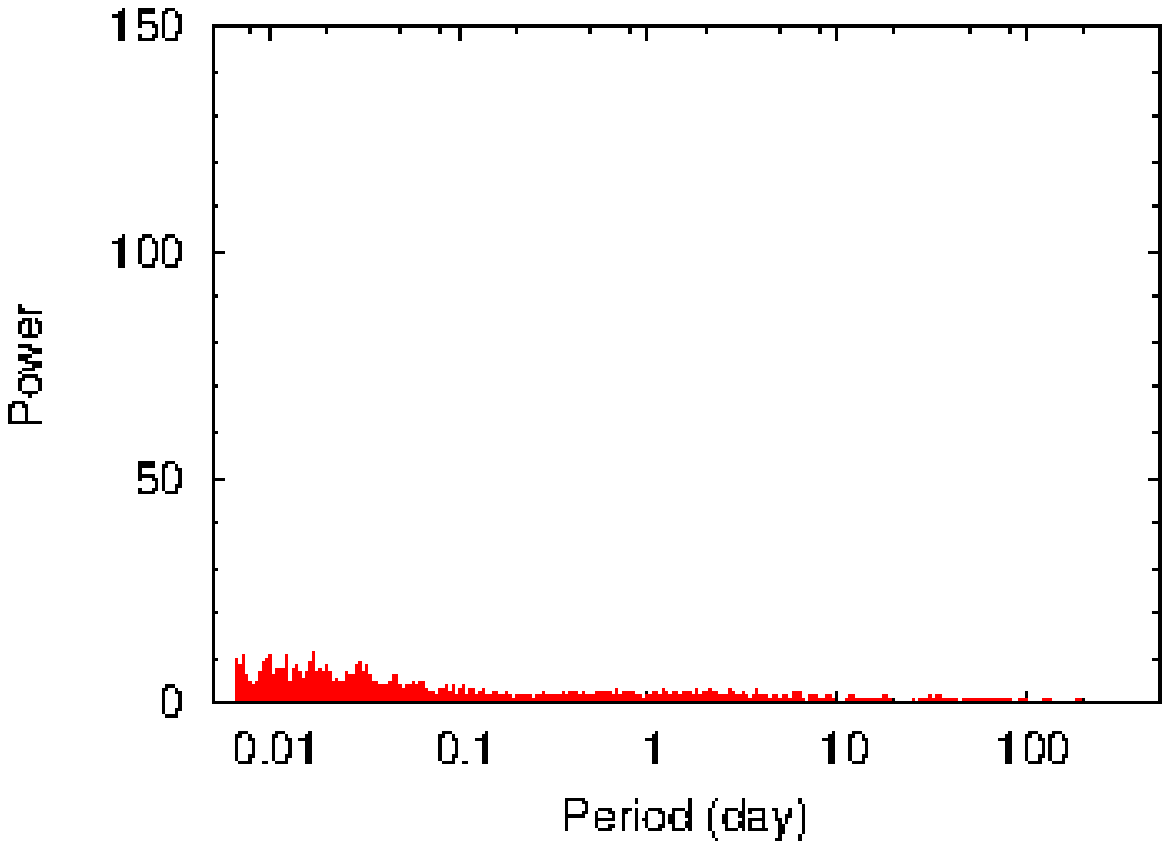}
\caption{Upper panel: Lomb-Scargle power spectrum for the star V56. The highest peak at 0.173 day is taken as the estimated 
period. Lower panel: Power spectrum of the same star after randomizing its amplitude.}
 \label{fig : PS}
\end{figure}

\begin{figure}
\centering
 \includegraphics[width=0.97\columnwidth, angle=0]{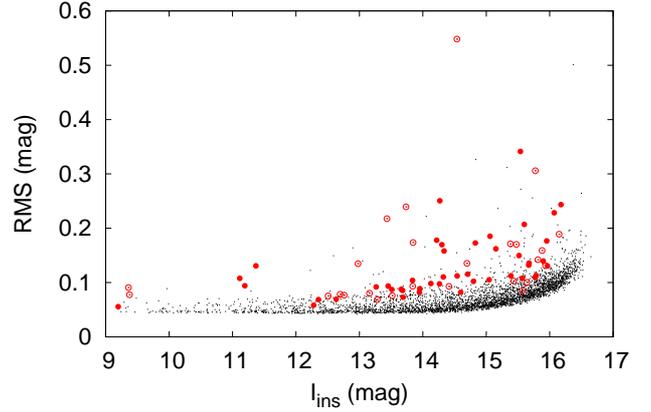}
 \caption{The RMS dispersion of instrumental magnitude as a function of mean instrumental magnitude
of the stars in the Sh 2-170 region (black dots). The open and filled circles indicate 
non-periodic and periodic variable candidates, respectively.}
 \label{fig : RMS}
\end{figure}

\subsection{Identification of variables}
\label{variable identification}

Differential photometry have been used to identify variables in  
$18^{\prime}.4 \times 18^{\prime}.4$ FOV of the Sh 2-170 region. This procedure automatically cancels out the other interfering 
effects such as changes in photon counts due to sky variation, effects of airmass, instrumental signatures etc. Briefly, we 
divided all the stars in the frame in different magnitude bins (e.g., 10-11, 11-12, 12-13 mag and so on), and in each magnitude 
bin we identified a comparison star to generate differential LCs of our target sources. For this we created all the possible pairs 
of stars in a particular magnitude bin and calculated the difference 
between the magnitudes of the pair stars for all the observed frames.  One of the 
star of the pair having lowest standard deviation for the difference in magnitude has been selected as the comparison star
for that magnitude bin. 
For better accuracy, we have used only those stars which are having photometric error better than 0.1 mag in $I_{c}$~band.  
We have also avoided those stars which are located in the nebulosity, near the bright star or at the  corners/edges of the CCDs.

The $I_{c}$ band differential magnitude ($\Delta m$) in the sense `$target-comparison$' were plotted as a function of 
Julian date to generate LCs of the target stars 
and to identify variables. 
In the first step, we visually checked all the LCs for any variability. 
Any star which exhibited a systematic 
visual variation larger than the scatter in comparison star 
is considered as a variable star. 
There were some stars which do not show any intra-night variability
but exhibit night-night variation.
In the second step, we  used the 
 \textit{Period\footnote{http://www.starlink.rl.ac.uk/docs/sun167.htx/sun167.html}}
software based on the Lomb-Scargle (LS) periodogram \citep{1976Ap&SS..39..447L,1982ApJ...263..835S}
to determine the periods of all the stars and phase-folded  the LCs 
to identify the periodic variables visually. 
The LS method is effective even in the case of unevenly sampled data sets which are common 
in a majority of the astronomical observations. To check 
whether the data gaps could produce any false periodicity, we performed LS periodogram  
analysis on the  periodic star after randomizing its amplitude
by using linux command-line utility i.e., shuf\footnote{http://www.gnu.org/software/coreutils/shuf}.
None of the power spectra of randomized LC of identified periodic variables show any periodic signature. 
As an example,  power spectra of the periodic variable V56 and its randomized LC are shown in 
Fig. \ref {fig : PS}. The maximum power is found at 0.173 day for this periodic variable 
whereas no signature of periodicity is visible in the randomized LC. 
We have also verified the periods using the
\textit {NASA Exoplanet Archive Periodogram 
service\footnote{https://exoplanetarchive.ipac.caltech.edu/cgi-bin/Pgram/nph-pgram}} 
and {PERIOD04\footnote{http://www.univie.ac.at/tops/Period04}  
\citep{2005CoAst.146...53L}. The periods obtained using these programs generally matched well.
Once the periodic variables were identified, 
the remaining LCs were once more checked visually to identify non-periodic variables.
}

The RMS dispersion of the
magnitudes as a function of mean instrumental magnitude for all the target stars is shown in 
Fig. \ref{fig : RMS}. As expected the dispersion 
increases towards the fainter end. Identified variables are shown with filled (periodic)  and open (non-periodic) 
circles. Some of the stars with a very high rms in Fig \ref{fig : RMS} were not designated as 
variables due to unusual high photometric errors
(possible reason may be bad pixels, bright background of the nearby star,
residual from the cosmic corrections, etc) as compared to stars 
in the same magnitude bin or large deviation in magnitude estimation.
These stars were further checked visually to ascertain the variability.

\begin{figure}
\centering
\includegraphics[width=1.0\columnwidth, angle= 0]{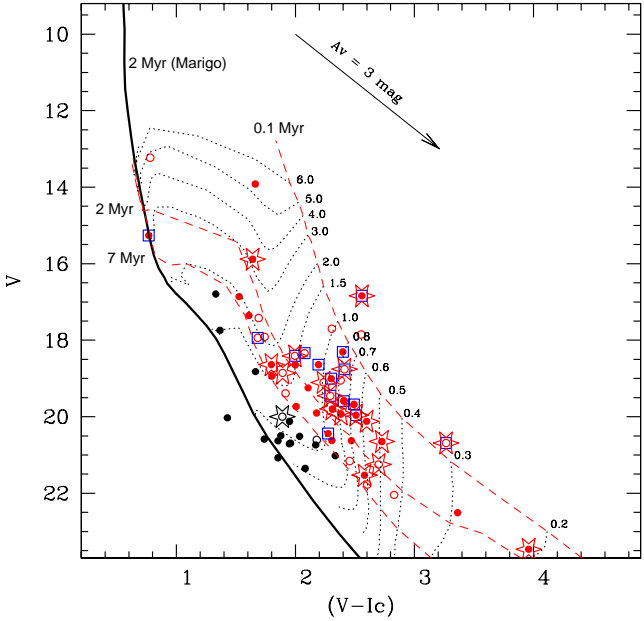}
\caption{ $V/(V-I_{c})$ CMD for the variable sources in the Sh 2-170 region. Filled and open circles represent 
periodic and non-periodic variables, respectively. Red and black colours represent member (age $\leq$ 7 Myr) 
and  non-member variables, respectively. 
Blue open squares represent variable stars with membership probability higher 
than 80\% as derived from the proper PM data. Class\,{\sc ii} YSOs are plotted with star symbols. 
The post-MS isochrone and the PMS isochrones are the same as in Fig. \ref{fig : MY_CMD}. The evolutionary 
tracks for various masses by \citet{2000A&A...358..593S} have 
also been plotted with black dotted curves. The arrow indicates the reddening vector.} 
 \label{fig : V_CMD}
\end{figure}

We  identified 47 and 24 stars as  periodic and non-periodic variables. 
The identification number, coordinates, period and other parameters of the identified variables
are listed in Table \ref{Variables}. The periods and amplitude of the variables range between  4 hrs and 18 
days and from 0.1 to 2.0 magnitude, respectively. 
 The optical $V/(V-I_c)$ CMD as discussed in Section \ref{distace and age of cluster} will further be used to 
classify the identified variables as PMS members or field stars (cf. next section).
Therefore, the LCs of most of the variables (66) will be discussed in ensuing sections except for
five having no $V$ band detection, i.e., V41, V42, V66, V68 and V71. 
The LCs of these variables are shown in Appendix A. 
For the two variables (V26 and V43; cf. Table \ref{Variables}) without $V$~band photometry,
we transformed their available $SDSS$ magnitudes (g, r, i) to $V$ and $I_{c}$ 
magnitudes using the available transformation 
equations\footnote{https://www.sdss.org/dr12/algorithms/sdssubvritransform/}. 

\begin{figure}
\centering
\includegraphics[width=0.7\columnwidth, angle= 0]{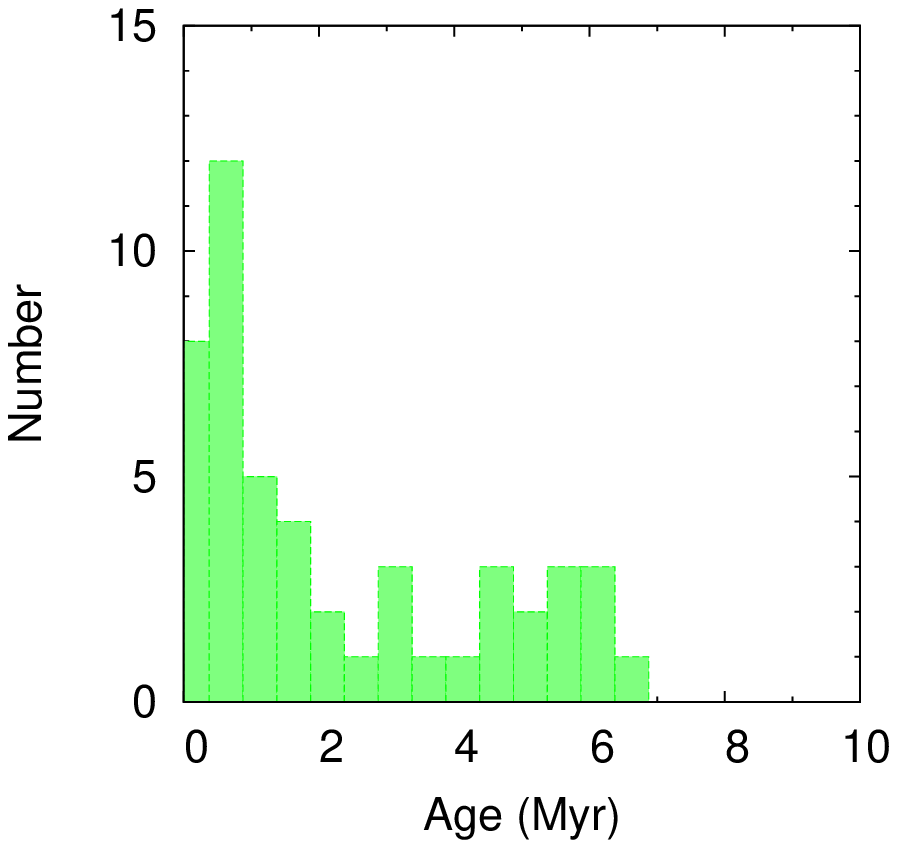}
\includegraphics[width=0.7\columnwidth, angle= 0]{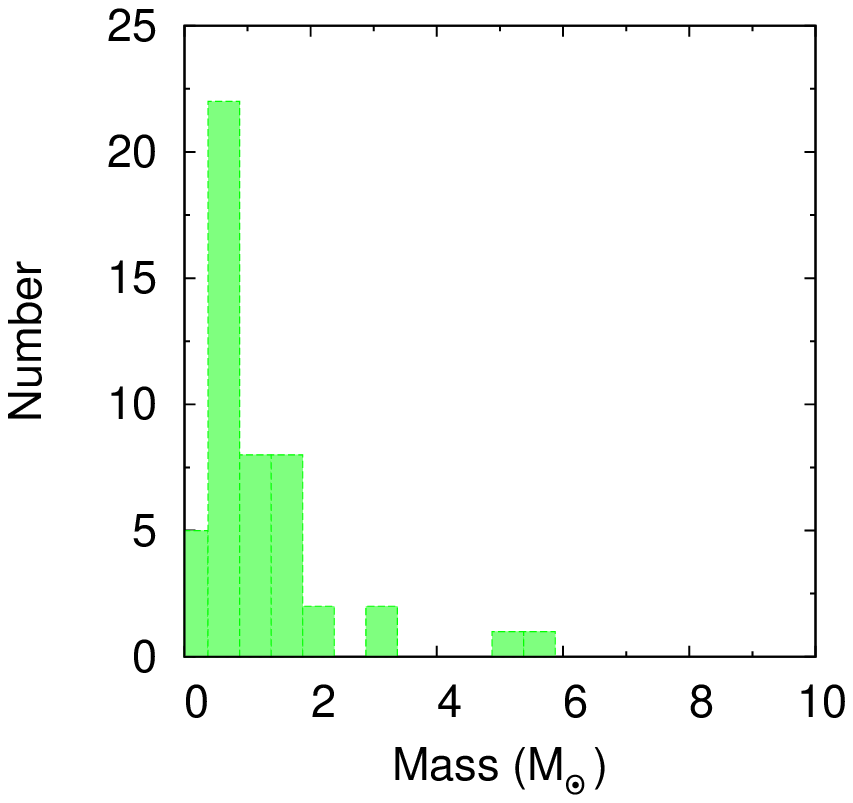}
\caption{Age (upper panel) and mass (lower panel) frequency distribution for PMS variable stars.}
 \label{fig: Hist_age_mass}
\end{figure}

\begin{figure}
\centering
\includegraphics[width=0.8\columnwidth, angle= 0]{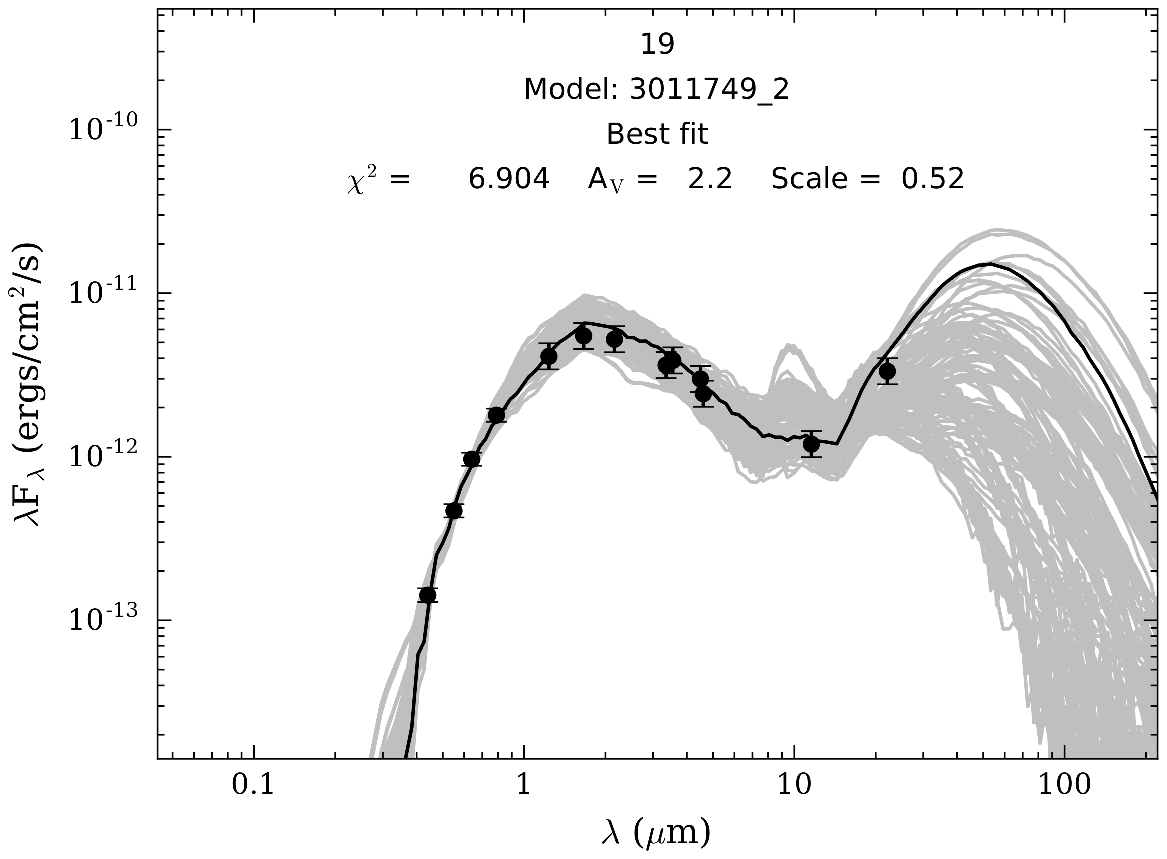}
\includegraphics[width=0.8\columnwidth, angle= 0]{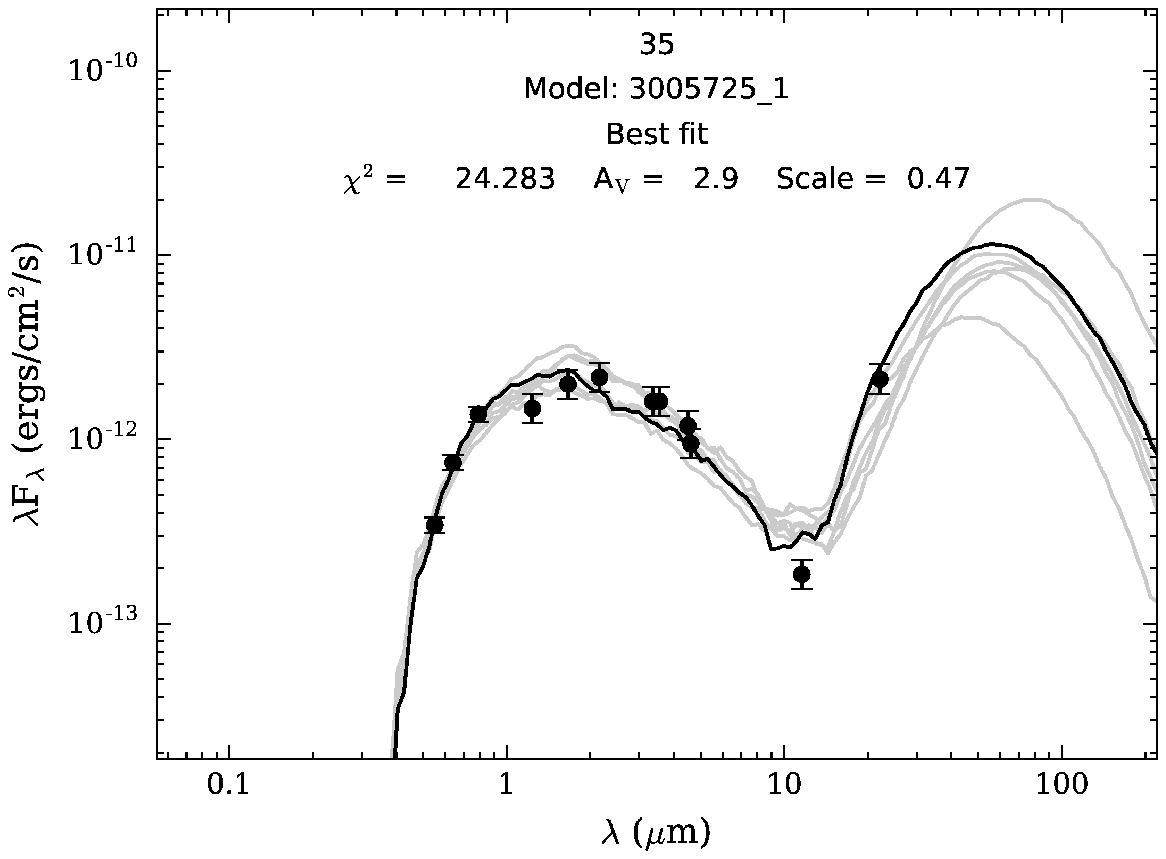}
\caption{Sample SEDs of two sources in the Sh 2-170 region created by the SED fitting tools of 
\citet{2007ApJS..169..328R}. The black curve shows the best fit and the grey curves show the subsequent 
well fits. The filled circles with error bars denote the input flux values. }
 \label{fig:SED}
\end{figure}

\subsection{Determination of physical parameters of the variables}

\subsubsection{Through HR diagram}
\label{variable_CMD}

 Fig. \ref{fig : V_CMD} shows  $V/(V-I_{c})$ CMD for 66 variables where
 open and filled circles denote non-periodic and periodic variables, respectively.  
The post-MS isochrone (black thick curve) and the PMS isochrones are the same as in 
Fig. \ref{fig : MY_CMD}. The evolutionary tracks of different masses (black dotted curves) 
by \citet{2000A&A...358..593S} are also shown. All the isochrones and evolutionary tracks are corrected for 
the distance and reddening (cf. Section \ref{distace and age of cluster}). Seventeen of the variables show 
excess IR emission, whereas fifteen of the variables are cluster members as determined from membership 
probability. These are shown with red star and blue square symbols, respectively. 
Most of these variables have ages $\le$ 2 Myr with an upper age limit of 7 Myr,
as discussed in Section \ref{distace and age of cluster}.
Therefore, 49 variables (29 periodic) having age $\leq 7$ Myr are considered to be PMS stars
associated with Sh 2-170  (red symbols). 
Remaining 17 variables (15 periodic)  are considered as non-member MS/field stars (black symbols).

The individual age and mass of each PMS variables can be determined from 
their position in the CMD by using the procedure discussed by \citet{2009MNRAS.396..964C} 
and \citet{2017MNRAS.467.2943S}. 
The estimated age and mass of the PMS variables and the YSOs are 
given in  Table \ref{Variables} and Table \ref{YSO}, respectively. The age and mass distributions of the PMS variables are shown 
in Fig. \ref{fig: Hist_age_mass}. The age distribution shows a peak around 1 Myr and an age spread of up to 7 Myr.  
Average age and mass of the 49 PMS variables associated with Sh 2-170 are found to be 2.4$\pm$0.6 
Myr and 1.3$\pm$0.1 M$_\odot$, respectively.

\begin{figure*}
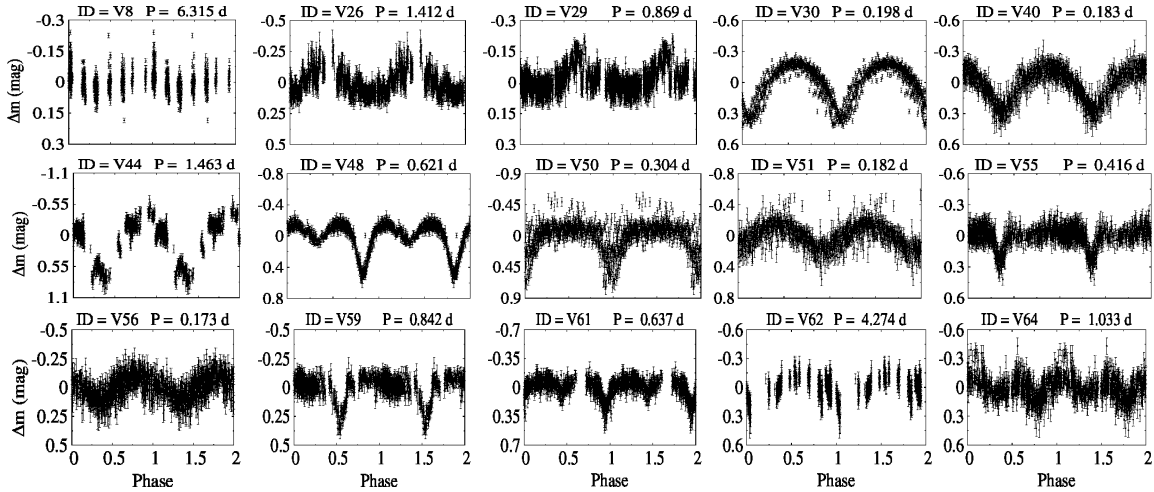

\centering
\includegraphics[width = 1.9 cm, height = 3.0 cm, angle= 270]{figures/periodic/ms/008pha.lreps}
\hspace{0.03 cm}
\includegraphics[width = 1.9 cm, height = 2.8 cm, angle= 270]{figures/periodic/ms/026pha.lreps}
\hspace{0.03 cm}
\includegraphics[width = 1.9 cm, height = 2.8 cm, angle= 270]{figures/periodic/ms/029pha.lreps}
\hspace{0.03 cm} 
\includegraphics[width = 1.9 cm, height = 2.8 cm, angle= 270]{figures/periodic/ms/030pha.lreps} 
\vspace{0.1cm}
\includegraphics[width = 1.9 cm, height = 2.8 cm, angle= 270]{figures/periodic/ms/040pha.lreps} \\
\hspace{0.03 cm} 
\includegraphics[width = 1.9 cm, height = 3.0 cm, angle= 270]{figures/periodic/ms/044pha.lreps}
\hspace{0.03 cm}
\includegraphics[width = 1.9 cm, height = 2.8 cm, angle= 270]{figures/periodic/ms/048pha.lreps}
\hspace{0.03 cm} 
\includegraphics[width = 1.9 cm, height = 2.8 cm, angle= 270]{figures/periodic/ms/050pha.lreps}
\vspace{0.1cm}
\includegraphics[width = 1.9 cm, height = 2.8 cm, angle= 270]{figures/periodic/ms/051pha.lreps}
\hspace{0.03 cm} 
\includegraphics[width = 1.9 cm, height = 2.8 cm, angle= 270]{figures/periodic/ms/055pha.lreps} \\
\hspace{0.03 cm} 
\includegraphics[width = 2.3 cm, height = 3.0 cm, angle= 270]{figures/periodic/ms/056pha.lreps} 
\hspace{0.03 cm}
\includegraphics[width = 2.3 cm, height = 2.8 cm, angle= 270]{figures/periodic/ms/059pha.lreps} 
\hspace{0.03 cm}
\includegraphics[width = 2.3 cm, height = 2.8 cm, angle= 270]{figures/periodic/ms/061pha.lreps}
\hspace{0.03 cm}
\includegraphics[width = 2.3 cm, height = 2.8 cm, angle= 270]{figures/periodic/ms/062pha.lreps} 
\vspace{0.1cm}
\includegraphics[width = 2.3 cm, height = 2.8 cm, angle= 270]{figures/periodic/ms/064pha.lreps}
\caption{Phase folded differential LCs of the 15 MS/field periodic variables. 
The identification numbers and periods (days) of the stars are given on the top of each panel.}
 \label{fig: LC_MS_P}
\end{figure*}

\subsubsection{Through spectral energy distribution}
\label{SED}

We constructed SEDs of the PMS variables  using the grid models and fitting tools of
\citet{2003ApJ...598.1079W, 2003ApJ...591.1049W, 2004ApJ...617.1177W} and
\citet{2006ApJS..167..256R, 2007ApJS..169..328R}
to characterise and understand their nature.
This method has been extensively used in our previous studies \citep[see. e.g.,][and references therein]{2016ApJ...822...49J,2017MNRAS.467.2943S}.
We were able to construct SEDs for the 43 PMS variables using the available 
optical, NIR (2MASS) and MIR (WISE, $Spitzer$) data, with a condition 
that each star should  have photometric data at least in five bands.  
The SED fitting tool fits each of the models to the data allowing the distance and extinction 
as free parameters. The distance for
the region is taken as $3.1\pm0.2$ kpc (cf. Section \ref{distace and age of cluster}) and we varied $A_{V}$ in a range 2.2 to 30 mag 
\citep{2012NewA...17..160B}. We further set photometric uncertainties of 10\% for the optical 
and 20\% for both the NIR and MIR data. These values are adopted instead of the formal errors in the catalog in order to fit 
without any possible biases caused by underestimating the flux uncertainties. We obtained the physical parameters of the PMS 
variables using the relative probability distribution for the stages of all the `well-fit' models. The well-fit models for each source 
are defined by $\chi^{2}- \chi^{2}_{min}$ $\leq$ 2$N_{data}$, where $\chi^{2}_{min}$ is the goodness-of-fit parameter for the 
best-fit model and $N_{data}$ is the number of input data points. In Fig. \ref{fig:SED}, we show example SEDs for the two PMS 
variables, where the solid black curves represent the best-fit and the grey curves 
are the subsequent well-fits. Table \ref{Variables} and Table \ref{YSO}
list the age, mass and other relevant parameters of the PMS variables and YSOs estimated from the SED analysis. 
The average age and mass of the  43 PMS variables are found to be $3.2\pm1.8$ Myr and $2.6\pm0.8$ M$_\odot$,  respectively. 
These values are comparable within errors to those derived from the CMD (age=2.4$\pm$0.6 Myr, mass=1.3$\pm$0.1 M$_\odot$).

\begin{figure*}
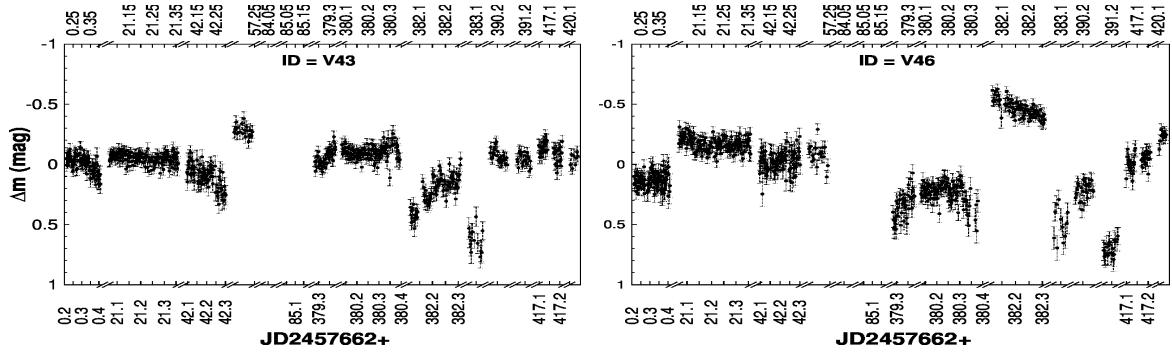

\centering
\includegraphics[width= 4.5 cm,height = 7.5 cm, angle=270]{figures/non-periodic/MS/043-np.lreps}
\hspace{0.05 cm}
\includegraphics[width= 4.5 cm,height = 7.5 cm, angle=270]{figures/non-periodic/MS/046-np.lreps} \\
\vspace{0.2 cm}
\caption{The differential LCs of the 2 non-periodic MS/field stars.  } 
\label{fig: LC_MS_NP}
\end{figure*}

\begin{figure}
\centering
\includegraphics[width=0.30\textwidth, angle = 270]{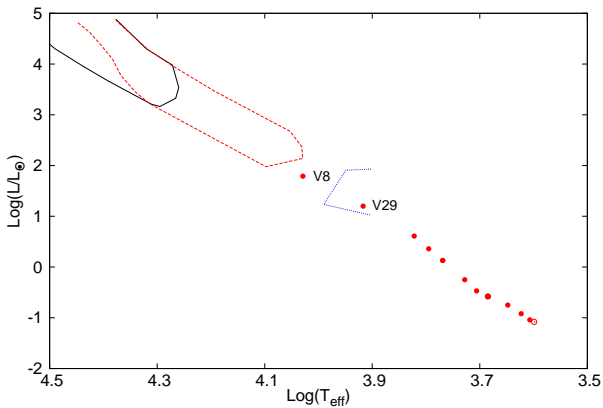}
 \caption{ Bolometric luminosity vs effective temperature 
of the MS/field star variables in the Sh 2-170 region. The regions with black continuous, red dashed and blue dotted borders indicate 
those for $\beta$ Cephi, SPB and $\delta$ Scuti stars, respectively. The positions of V29 and V8 in the HR diagram indicate that 
they are $\delta$ Scuti and new type of variable stars, respectively. Filled and open circles represent periodic and non-periodic 
variables, respectively.}
\label{fig : Teff_Bol} 
\end{figure}

\subsection{Classification of the identified variable stars}

The association of the 71 identified  variables  with Sh 2-170 has been checked  
on the basis of their membership probability (cf. Section \ref{membership}), 
the position in the CMD (cf. Section \ref{variable_CMD}), and the presence or absence of excess IR emission  
(cf. Section \ref{distace and age of cluster}),
and the details about their classification and nature of variability are given in Table \ref{variability nature}.   
Periodic/non-periodic nature of the variables are also mentioned in the table. 
Five of the identified variables could not be classified due to non-availability of their $V$ band data. Out of the remaining 
66 variables, 44 (66\%) exhibit periodicity. Forty nine variables are found to be PMS sources, 
of which 29 (59\%) variables are periodic. 
Most of the PMS variables have age and mass in the  range 
of  $\sim$ 0.1 - 2 Myr and $\sim$ 0.2 - 3 M$_\odot$, indicating that they are most probably T-Tauri stars.  
Seventeen of them exhibit excess IR emission and are
classified as Class\,{\sc ii} sources  (cf. Section \ref{distace and age of cluster}).
The remaining 32 PMS variable stars having either insignificant or no IR-excess,
may belong to the Class\,{\sc iii} category. Seventeen stars are found to be MS/field stars
and 15 (88\%) of them have periodic LCs.

\subsubsection{MS/field variables}

The variability in MS population is mainly caused by pulsation. 
The MS variability of various kind of pulsators  like $\beta$ Cep, $\delta$ Scuti, 
SPB, $\gamma$ Doradus etc. has been extensively studied
\citep[e.g.,][]{1997MNRAS.289...25B,2011MNRAS.417..591B,2011A&A...533A..70H,2013A&A...554A.108M}. 
A new class of variable stars situated in the HR diagram between the red end of SPB and the blue end of $\delta$ Scuti,  
where stars are not expected to occur according to the 
classical stellar models, are also reported \citep[see e.g.,][]{2013A&A...554A.108M,2014MNRAS.442..273L}. 

We have identified 17 MS/field variables and 15 of them are periodic having periods in the 
range $\sim$4 hrs - 6 days and $I_{c}$ band amplitudes of $\sim$0.35 - 1.58 mag.
Their respective phase-folded LCs are shown in Fig. \ref{fig: LC_MS_P}.
A higher percentage of periodic variables in the MS/field stars 
is natural as their variability is dominated by their pulsating behaviour \citep{1997MNRAS.289...25B,2011MNRAS.417..591B}.
The LCs of V48, V61 and V64 clearly show two dips which
resemble to those of $\beta$ Lyrae type binary stars.
The $\beta$ Lyrae type systems are composed of two stars typically with different evolutionary states. 
The binaries are in a tight orbit and mass transfer can take place in semi-detached systems \citep{2008AJ....136.1067H}.
V48 and V61 show similar periods of about 15 hrs with amplitudes of 0.87 and 0.7 mag, respectively. V64 shows a little 
longer period of about 1 day with an amplitude of 0.85 mag.
The $\beta$ Lyrae type variables generally have periods longer than 1 day \citep{2008AJ....136.1067H} with few exceptions
(e.g., 0.29 days for HD 105575 and 198.5 days for HD 105998\footnote{http://www.sai.msu.su/gcvs/cgi-bin/search.cgi?search=W+Cru}).
The secondary dip  (0.39 mag) in V48 is less than half of the primary dip (0.87 mag).
A small `bump' could also be seen around the secondary dip. In Fig. \ref{fig: LC_MS_NP}. We also show the
remaining LCs of 2 non-periodic MS/field stars showing variations in the range $\sim$1.1 - 1.4 mag. 

To further  characterise these MS/field variables, we plot in
Fig. \ref{fig : Teff_Bol} their effective
 temperature ($T_{eff}$) versus bolometric luminosity (L/L$_\odot$) 
HR diagram. The absolute magnitudes were estimated by using the distance given by \citet{2018AJ....156...58B} and 
\citet{2018A&A...616A...1G,2016A&A...595A...1G} and assuming the normal extinction law 
\citep{1980AJ.....85...17M,1969ApJ...157..611G}. 
The absolute magnitudes of these stars are then matched with those given in the  theoretical MS by 
\citet{2013ApJS..208....9P}\footnote{http://www.pas.rochester.edu/\\$\sim$emamajek/EEM\_dwarf\_UBVIJHK\_colors\_Teff.txt}  and 
the corresponding values of $T_{eff}$ and L/L$_\odot$ are taken. 
The associated errors could be due to uncertainty in theoretical models, parallax values and/or  in the extinction values. 
Fig. \ref{fig : Teff_Bol}  indicates that V29 is a $\delta$ Scuti  star (blue dotted region). 
The amplitude and period of this star are 0.43 mag and $\sim$20 hrs,  respectively. 
V8 lies in between the SPB and $\delta$ Scuti regions. This one could be a variable of a new class. 
Its amplitude in $I_{c}$~band and period are 0.35 mag and 6.3 days, respectively. 
The remaining stars are of lower surface temperature/mass (9000K/$\sim$2M$_\odot$ - 4000K/$\sim$0.6M$_\odot$). 
Detailed analysis of these MS/field variables are beyond the scope of this study in which we try to concentrate mainly on 
variability of PMS stars, as discussed in detail in the next section. 

\begin{figure*}
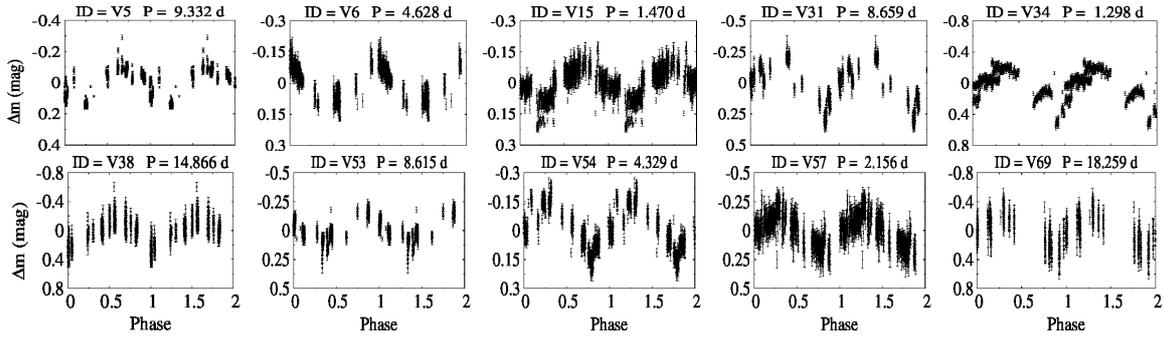

\centering
\includegraphics[width = 1.9 cm, height = 3.0 cm, angle= 270]{figures/periodic/class-II/005pha.lreps}
\hspace{0.03 cm}
\includegraphics[width = 1.9 cm, height = 2.8 cm, angle= 270]{figures/periodic/class-II/006pha.lreps}
\hspace{0.03 cm}
\includegraphics[width = 1.9 cm, height = 2.8 cm, angle= 270]{figures/periodic/class-II/015pha.lreps} 
\hspace{0.03 cm}
\includegraphics[width = 1.9 cm, height = 2.8 cm, angle= 270]{figures/periodic/class-II/031pha.lreps} 
\hspace{0.03 cm}
\includegraphics[width = 1.9 cm, height = 2.8 cm, angle= 270]{figures/periodic/class-II/034pha.lreps} \\
\vspace{0.05 cm}
\hspace{0.02 cm}
\includegraphics[width = 2.3 cm, height = 3.0 cm, angle= 270]{figures/periodic/class-II/038pha.lreps}
\hspace{0.03 cm}
\includegraphics[width = 2.3 cm, height = 2.8 cm, angle= 270]{figures/periodic/class-II/053pha.lreps}
\hspace{0.03 cm}
\includegraphics[width = 2.3 cm, height = 2.8 cm, angle= 270]{figures/periodic/class-II/054pha.lreps}
\hspace{0.03 cm}
\includegraphics[width = 2.3 cm, height = 2.8 cm, angle= 270]{figures/periodic/class-II/057pha.lreps}
\hspace{0.03 cm}
\includegraphics[width = 2.3 cm, height = 2.8 cm, angle= 270]{figures/periodic/class-II/069pha.lreps}

\caption{Phase folded differential LCs of 10 periodic  
Class\,{\sc ii} variables. 
The identification numbers and periods (days) of the stars are given on the top of
each panel.}
 \label{fig: LC_Class_II_P}
\end{figure*}

\begin{figure*}
\centering
\hspace{-0.1 cm}
\includegraphics[width = 1.9 cm, height = 3.2 cm, angle= 270]{figures/periodic/class-III/001pha.lreps}
\hspace{0.03 cm}
\includegraphics[width = 1.9 cm, height = 2.8 cm, angle= 270]{figures/periodic/class-III/004pha.lreps}
\hspace{0.03 cm}
\includegraphics[width = 1.9 cm, height = 2.8 cm, angle= 270]{figures/periodic/class-III/009pha.lreps}
\hspace{0.03 cm}
\includegraphics[width = 1.9 cm, height = 2.9 cm, angle= 270]{figures/periodic/class-III/010pha.lreps} 
\hspace{0.03 cm}
\includegraphics[width = 1.9 cm, height = 2.9 cm, angle= 270]{figures/periodic/class-III/016pha.lreps} \\
\vspace{0.05 cm}
\hspace{-0.2 cm}
\includegraphics[width = 1.9 cm, height = 3.3 cm, angle= 270]{figures/periodic/class-III/017pha.lreps}
\hspace{-0.1 cm}
\includegraphics[width = 1.9 cm, height = 2.9 cm, angle= 270]{figures/periodic/class-III/018pha.lreps}
\hspace{-0.1 cm}
\includegraphics[width = 1.9 cm, height = 2.9 cm, angle= 270]{figures/periodic/class-III/021pha.lreps} 
\hspace{0.1 cm}
\includegraphics[width = 1.9 cm, height = 2.9 cm, angle= 270]{figures/periodic/class-III/023pha.lreps}
\hspace{0.1 cm}
\includegraphics[width = 1.9 cm, height = 2.8 cm, angle= 270]{figures/periodic/class-III/024pha.lreps} \\
\vspace{0.05 cm}
\hspace{-0.2 cm}
\includegraphics[width = 1.9 cm, height = 3.3 cm, angle= 270]{figures/periodic/class-III/032pha.lreps}
\hspace{0.0 cm}
\includegraphics[width = 1.9 cm, height = 2.8 cm, angle= 270]{figures/periodic/class-III/033pha.lreps} 
\hspace{0.03 cm}
\includegraphics[width = 1.9 cm, height = 2.8 cm, angle= 270]{figures/periodic/class-III/037pha.lreps}
\hspace{0.03 cm}
\includegraphics[width = 1.9 cm, height = 2.9 cm, angle= 270]{figures/periodic/class-III/039pha.lreps}
\hspace{0.03 cm}
\includegraphics[width = 1.9 cm, height = 2.9 cm, angle= 270]{figures/periodic/class-III/045pha.lreps} \\
\vspace{0.05 cm}
\includegraphics[width = 2.3 cm, height = 3.2 cm, angle= 270]{figures/periodic/class-III/058pha.lreps} 
\vspace{0.1cm}
\includegraphics[width = 2.3 cm, height = 2.9 cm, angle= 270]{figures/periodic/class-III/060pha.lreps}
\hspace{0.03 cm}
\includegraphics[width = 2.3 cm, height = 2.9 cm, angle= 270]{figures/periodic/class-III/065pha.lreps}
\hspace{0.03 cm}
\includegraphics[width = 2.3 cm, height = 2.9 cm, angle= 270]{figures/periodic/class-III/067pha.lreps}

\caption{Phase folded differential LCs of 19 periodic Class\,{\sc iii} variables (WTTSs). 
The identification numbers and periods (days) of the stars are given on the top of
each panel.}
 \label{fig: LC_Class_III_P}
\end{figure*}

\begin{figure}
\centering
\includegraphics[width=0.35\textwidth]{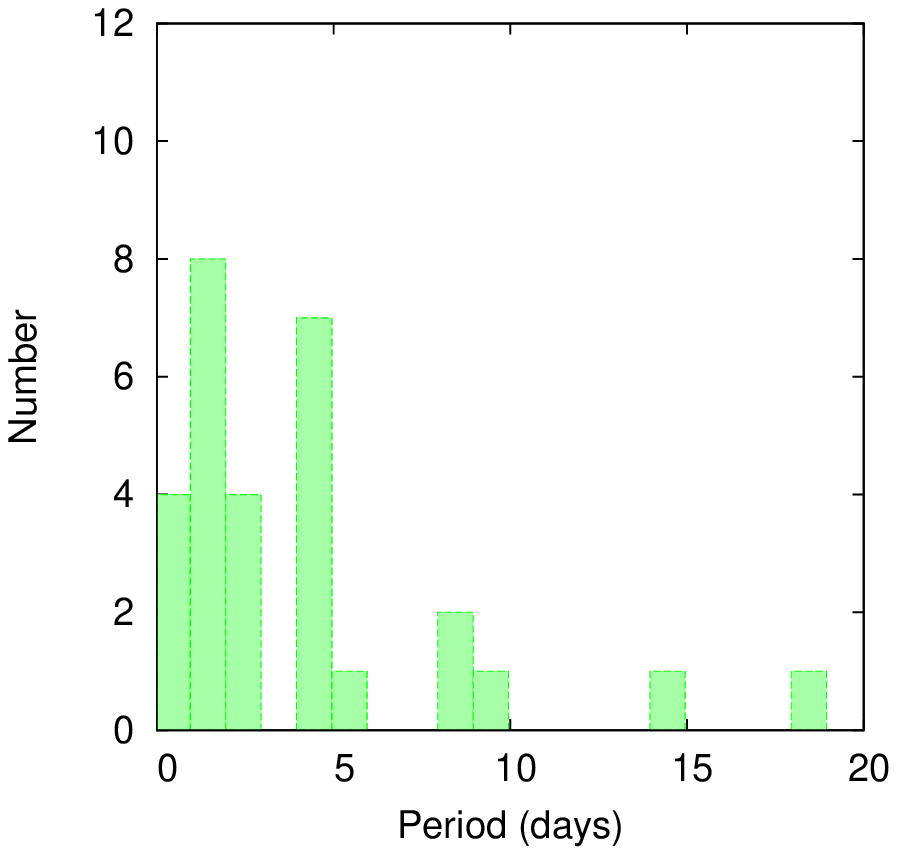}
\includegraphics[width=0.35\textwidth]{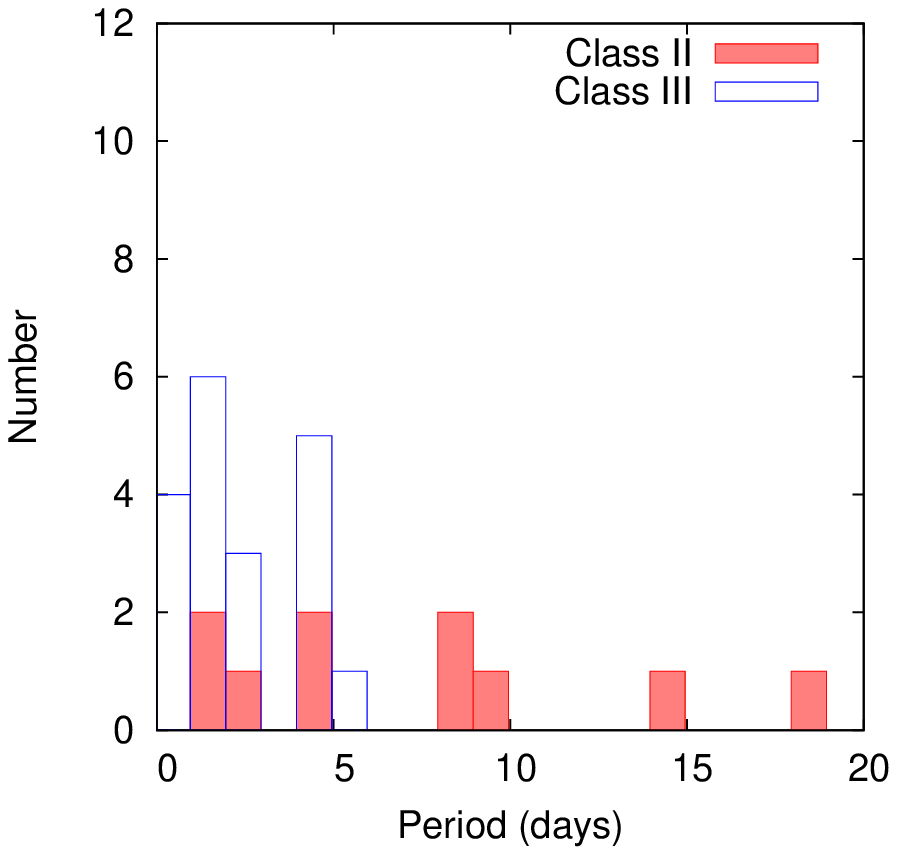}
 \caption{Top panel: Period distribution of PMS variables. Bottom panel: 
 Period distribution of Class\,{\sc ii} and Class\,{\sc iii} sources.}
 \label{fig : period distribution}
\end{figure}

\begin{figure*}
\centering
\includegraphics[width= 4.5 cm,height = 7.5 cm, angle=270]{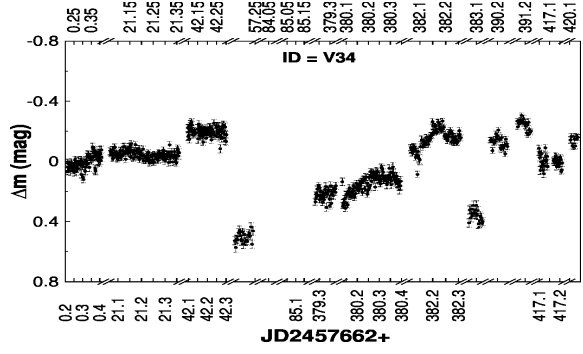} 
\caption{LC of V34.}
\label{fig: V34}
\end{figure*}

\section{DISCUSSION: PRE-main-sequence VARIABLES}
\label{sect:discussion}

The circumstellar disc plays a significant role in the photometric variation of both periodic and non-periodic PMS stars. Accretion 
and dust obscuration are responsible for most of the irregular variations in young stars \citep{1994AJ....108.1906H}.  Spot (both 
hot and cool) modulation is the most common cause of periodic variation. CTTSs have a thick disc, from which magnetically 
guided accretion give rise to strong $H\alpha$ emission. Erratic enhanced accretion causes hot spots on the stellar surface, 
which can lead to large amplitude variations. In WTTS the disc is more or less depleted and the accretion is inactive. Instead 
developed cool spots on the stellar surface emerges as the major modulator of variability. 

PMS stars with different masses have different amount of circumstellar disk material around them. Also they undergo 
mass-dependent evolution of the internal structure, which determines the dynamo process inducing stellar magnetic activities 
\citep{2009ARA&A..47..333D}. This not only influences the accretion but also controls the stellar rotation through disc-locking. 
In the disc-locking scenario \citep{1991ApJ...370L..39K}  
the disc-star interaction results in the transfer of angular momentum from the star to the circumstellar disc and consequently the 
star maintains almost a constant rotation rate until the coupling breaks due to the significant dissipation of 
the disc. The angular momentum could also be released through an enhanced stellar wind powered by the accretion of material 
from the disc  \citep{2005ApJ...632L.135M}.  In both the cases a correlation between rotation speed and presence of disc is 
expected in the sense that slow rotators must be surrounded by substantial circumstellar disc, whereas fast rotators should be in 
a process of disc dispersal. We discuss variability characteristics of the PMS stars in our sample in the subsequent  
subsections.

\begin{figure*}
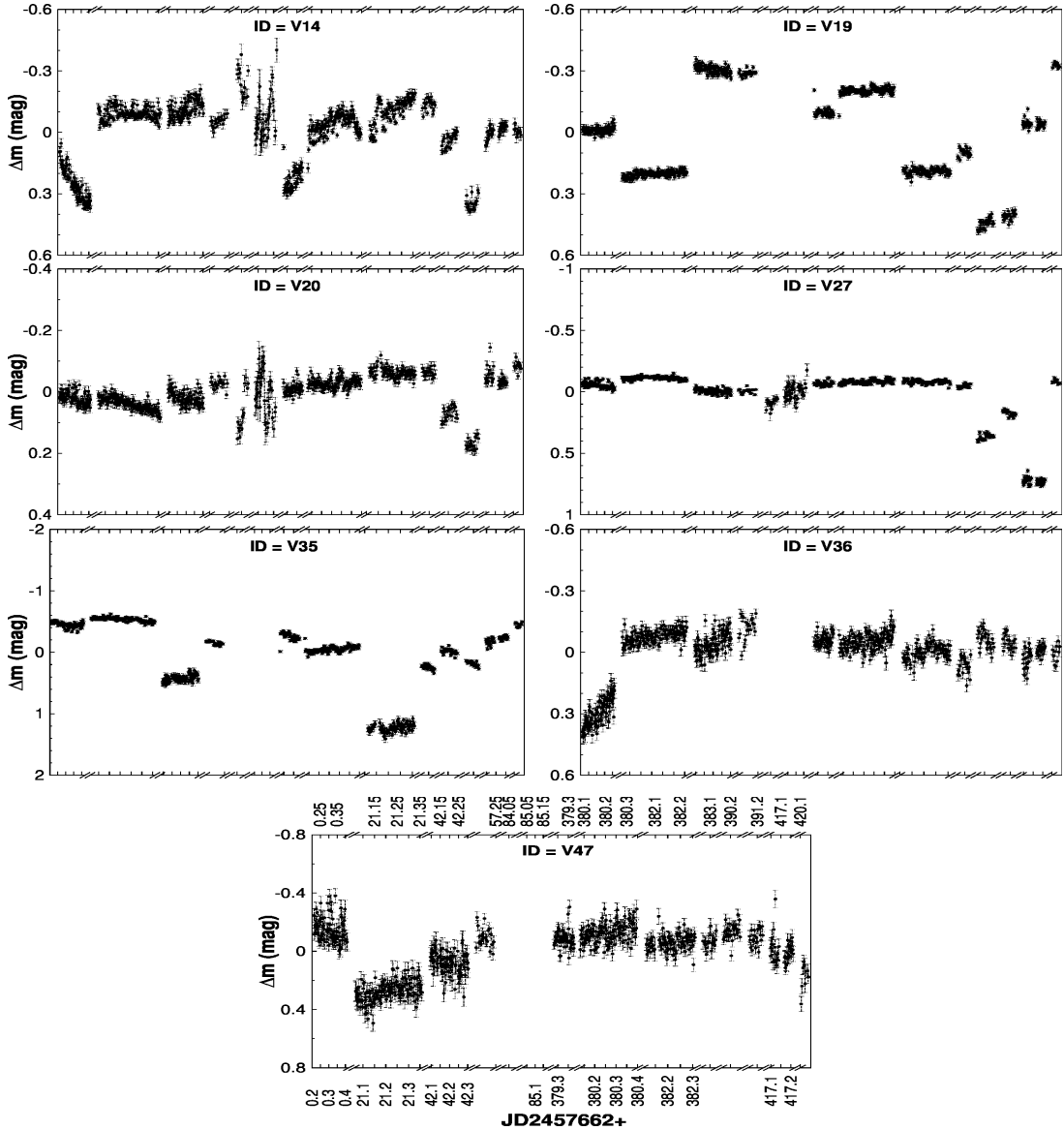

\centering
\includegraphics[width= 3.5 cm,height = 7.0 cm, angle=270]{figures/non-periodic/CTT/014-np.lreps}
\hspace{0.1 cm}
\includegraphics[width= 3.5 cm,height = 7.0 cm, angle=270]{figures/non-periodic/CTT/019-np.lreps} \\
\vspace{0.03 cm}
\includegraphics[width= 3.5 cm,height = 7.0 cm, angle=270]{figures/non-periodic/CTT/020-np.lreps}
\hspace{0.1 cm}
\includegraphics[width= 3.5 cm,height = 7.0 cm, angle=270]{figures/non-periodic/CTT/027-np.lreps} \\
\vspace{0.03 cm}
\includegraphics[width= 3.5 cm,height = 7.0 cm, angle=270]{figures/non-periodic/CTT/035-np.lreps}
\hspace{0.1 cm}
\includegraphics[width= 3.5 cm,height = 7.0 cm, angle=270]{figures/non-periodic/CTT/036-np.lreps} \\
\vspace{0.2 cm}
\includegraphics[width= 4.5 cm,height = 7.5 cm, angle=270]{figures/non-periodic/CTT/047-np.lreps}
\caption{Differential LCs of 7  non-periodic Class\,{\sc ii} sources (CTTSs). Identification number is given in each panel. 
The scale for X-axis is displayed in the case of V47 only and is the same for all the other variables.}
 \label{fig: LC_CTT}
\end{figure*}

\begin{figure*}
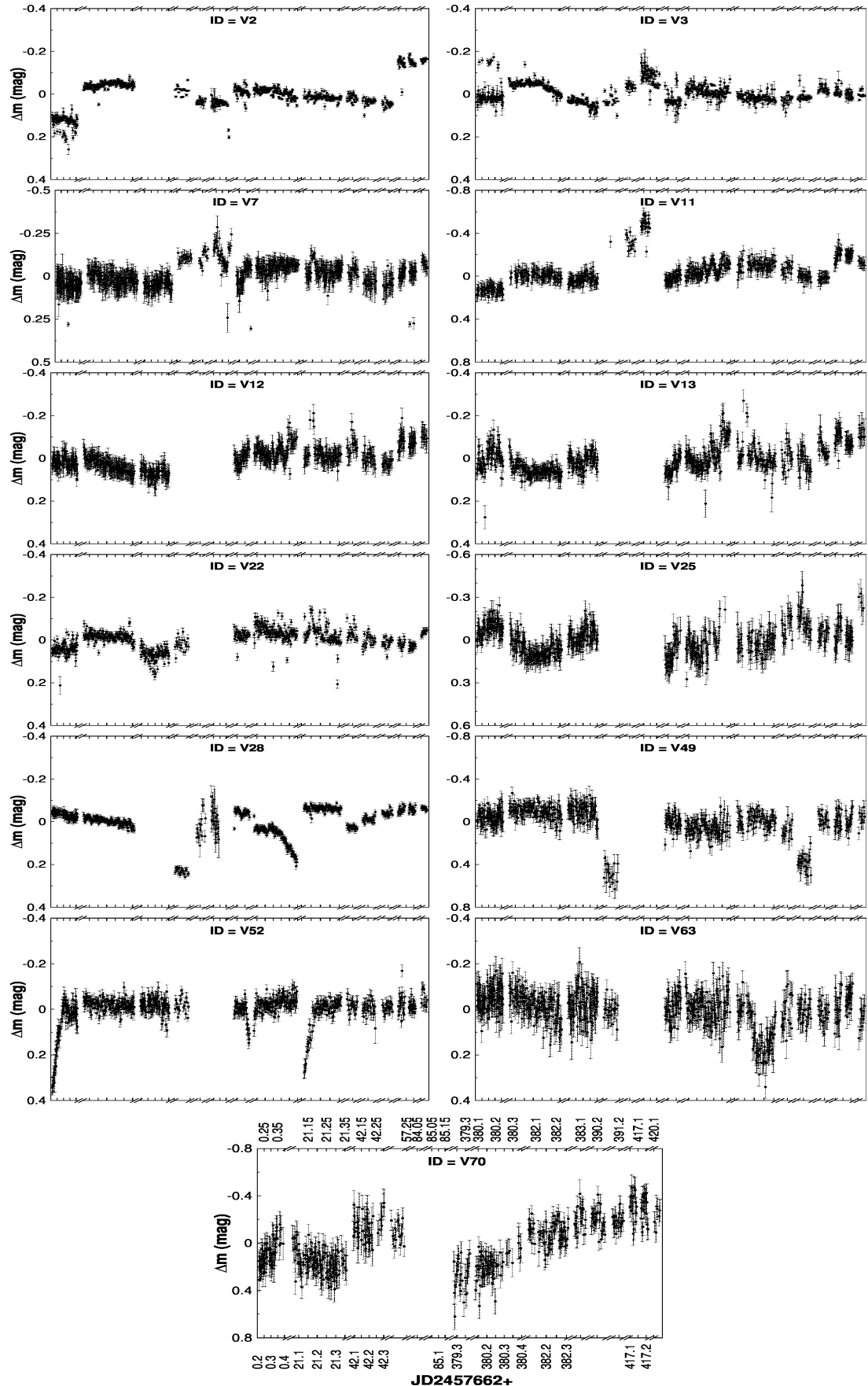

\centering
\includegraphics[width= 3.0 cm,height = 7.0 cm, angle=270]{figures/non-periodic/class_III/002-np.lreps}
\hspace{0.1 cm}
\includegraphics[width= 3.0 cm,height = 7.0 cm, angle=270]{figures/non-periodic/class_III/003-np.lreps} \\
\vspace{0.03 cm}
\includegraphics[width= 3.0 cm,height = 7.0 cm, angle=270]{figures/non-periodic/class_III/007-np.lreps}
\hspace{0.1 cm}
\includegraphics[width= 3.0 cm,height = 7.0 cm, angle=270]{figures/non-periodic/class_III/011-np.lreps} \\
\vspace{0.03 cm}
\includegraphics[width= 3.0 cm,height = 7.0 cm, angle=270]{figures/non-periodic/class_III/012-np.lreps} 
\hspace{0.1 cm}
\includegraphics[width= 3.0 cm,height = 7.0 cm, angle=270]{figures/non-periodic/class_III/013-np.lreps} \\
\vspace{0.03 cm}
\includegraphics[width= 3.0 cm,height = 7.0 cm, angle=270]{figures/non-periodic/class_III/022-np.lreps} 
\hspace{0.1 cm}
\includegraphics[width= 3.0 cm,height = 7.0 cm, angle=270]{figures/non-periodic/class_III/025-np.lreps} \\
\vspace{0.03 cm}
\includegraphics[width= 3.0 cm,height = 7.0 cm, angle=270]{figures/non-periodic/class_III/028-np.lreps} 
\hspace{0.1 cm}
\includegraphics[width= 3.0 cm,height = 7.0 cm, angle=270]{figures/non-periodic/class_III/049-np.lreps} \\
\vspace{0.03 cm}
\includegraphics[width= 3.2 cm,height = 7.0 cm, angle=270]{figures/non-periodic/class_III/052-np.lreps}
\hspace{0.1 cm}
\includegraphics[width= 3.2 cm,height = 7.0 cm, angle=270]{figures/non-periodic/class_III/063-np.lreps} \\
\vspace{0.2 cm}
\includegraphics[width= 4.5 cm,height = 7.5 cm, angle=270]{figures/non-periodic/class_III/070-np.lreps} 
\caption{Differential LCs of 13  non-periodic Class\,{\sc iii} variables. 
Identification number is given in each panel. The scale for X-axis is displayed in the case of V70 only and is the same for all the other variables.}
\label{Fig: LC_Class_III_NP}
\end{figure*}

\subsection{Periodic PMS variables: An insight into disc-locking}

PMS stars are supposed to release their angular momentum through different mechanisms 
as they make the transition from the protostellar state down to MS. 
Disc-locking is the most commonly invoked mechanism to regulate the angular momentum 
\citep[see e.g.,][]{2000AJ....120..349H}. \citet{2000AJ....120..349H,2002A&A...396..513H} have found a bimodal period distribution 
around 1 and 8 days for PMS variables with M > 0.25 M$_\odot$ in Orion Nebula Cluster (ONC). Several studies 
\citep[e.g.,][]{2000AJ....120..349H} explained this bimodal distribution through disc-locking mechanism. If a star is disc-locked further 
increase in its rotation speed due to contraction would be slowed down and a significant number of star will end up having similar 
rotation periods. After released from disc-locking, the stars will again increase their rotation speed. 

Twenty nine of our PMS variable stars show periodicity in their LCs. 
In Fig.  \ref{fig: LC_Class_II_P} and Fig. \ref{fig: LC_Class_III_P},  
we have plotted the phase folded LC of Class\,{\sc ii} (10) and Class\,{\sc iii} (19)
periodic variables, respectively. In general the periods and amplitudes of Class\,{\sc ii} variables are larger 
than Class\,{\sc iii} variables. All the Class\,{\sc iii} variables have period $<$ 6 days. 
As the variability signatures in WTTSs are dominated by the 
asymmetric distribution of cool spots on the stellar surface, 19 Class\,{\sc iii} sources showing  periodicity in the range  
of $\sim$4 hrs to 6 days with amplitudes of $\sim$0.15-0.68 mag 
(with an exception of V32 having an amplitude of 1.38 mag) are 
classified as WTTSs \citep[see also,][]{2019arXiv190600256B}. 
We show the period distribution for all 29 periodic PMS stars in 
the upper panel of Fig. \ref{fig : period distribution}. The distribution is bimodal 
with peaks near $\sim$1.5 and $\sim$4.5 days. 
We further subdivided this sample into Class\,{\sc ii} and Class\,{\sc iii} sources and 
their period distributions are also shown in the lower panel of Fig. \ref{fig : period distribution}. 
The bimodality is more clearly visible for Class\,{\sc iii} sources, 
whereas the period distribution for the Class\,{\sc ii} sources  is more or less flat.  
Here we would like to mention that uneven sampling and lack of observations 
for longer periods may be the reason for the small number of relatively slow rotators. 
\citet{2005A&A...430.1005L} have also found a similar bimodal period distribution 
for variables having $(R_{c}-I_{c})$ $>$ 1.3  (mass $\sim$0.25 M$_\odot$) with peaks at 1 and 4 days in NGC 2264.  
In the case of  IC 348, \citet{2005A&A...437..637L} have found a bimodal period distribution   around 3 and 8 days
similar to that seen in ONC for the stars having masses $>$ 0.25 M$_\odot$, but have
reported a statistically significant lack of rapidly rotating stars with respect to the ONC. 
They also concluded that in spite of the similarity in the ages and 
disc fractions between NGC 2264 and IC 348, the  marked difference in the period distributions 
between the two clusters presents a serious challenge to the disc-locking paradigm, 
as it provides no explanation for the difference. 
However, differences in the cluster environments and the physical conditions
leading to the star formation    
can lead to these observed differences in the period distribution \citep[see also, ][]{2005A&A...437..637L}.

Here, we would also like to mention the peculiar LC of V34 as shown in Fig. \ref{fig: V34}. In addition to the inter-night 
variability it also shows small intra-night variability. In the same night (within few hours) we see a small change in 
brightness in a periodic fashion. Multiple physical phenomena such as spot modulation, inner disc co-rotation and 
extinction events could be responsible for this.

\begin{figure}
\centering
\includegraphics[width=0.8\columnwidth, angle= 0]{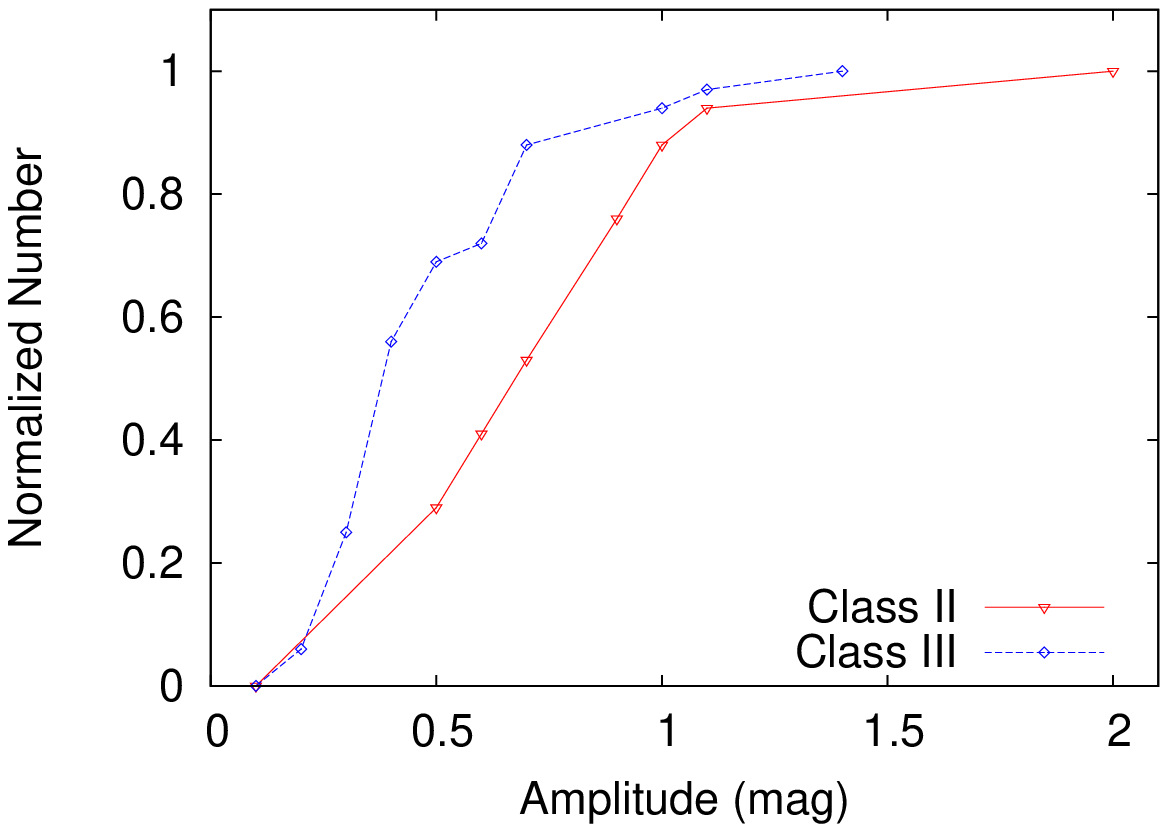}
\caption{Normalized cumulative amplitude distribution of Class\,{\sc ii} and Class\,{\sc iii} sources. }
 \label{fig: CAD}
\end{figure}

\subsection{Non-periodic PMS variables: Bursters and faders}

An increase in accretion rate from the circumstellar disc onto the star can give rise to a significant burst in magnitude (few mags)
that lasts over hours to days \citep{2018AJ....156...71C}. Similarly, variable circumstellar extinction can cause 
brightening/fading events of a few tenths of magnitude. They are of short phenomena of up to a few hours typically 
\citep[for e.g., AA Tau,][]{2018ApJ...852...56G}.

Twenty of our PMS variable stars do not show periodicity in their LCs. In Fig. \ref{fig: LC_CTT} and 
Fig. \ref{Fig: LC_Class_III_NP}, we have plotted the LCs of Class\,{\sc ii} (7) and Class\,{\sc iii} (13) non-periodic variables, 
respectively. Their amplitudes vary between 0.1 - 2 mag. 
The Class\,{\sc ii} variables show significantly large magnitude variations as compared to the Class\,{\sc iii} variables. 
Also, their variability features include either single or multiple fading/brightening events 
that last for different time-spans.These seven Class\,{\sc ii} variables are therefore classified as CTTSs 
\citep[see also,][]{2019arXiv190600256B}.
The photometric variations of these sources are found to be in between $I_c$$\sim$0.5 mag to $I_c$$\sim$2.0 mag. 
The source V35 shows a huge variation of 2 mag . On $11^{th}$ Nov 2016 it showed a decrease 
in magnitude from its semi-stable maximum. 
Again on $17^{th}$ Oct 2017 it showed a larger decrease, and  with 
a few fluctuations reached to its stable maximum brightness around $24^{th}$ Nov 2017. 
The source V27 is almost stable in most part of its LC with little fluctuation. 
After $18^{th}$ Oct 2017 it started fluctuating with a maximum magnitude decrease of around 
1 mag on $21^{th}$ Nov 2017. In the case of V14 and V36 we see significant intra-night 
fading and brightening events in few nights. We don't see any intra-night variation in V19, 
but inter-night variations are present in night to night with a maximum variation of 0.87 mag. 
The characteristics of variability found in these sources are typical of CTTSs.
 
\begin{figure*}
\centering
\includegraphics[width=0.35\textwidth]{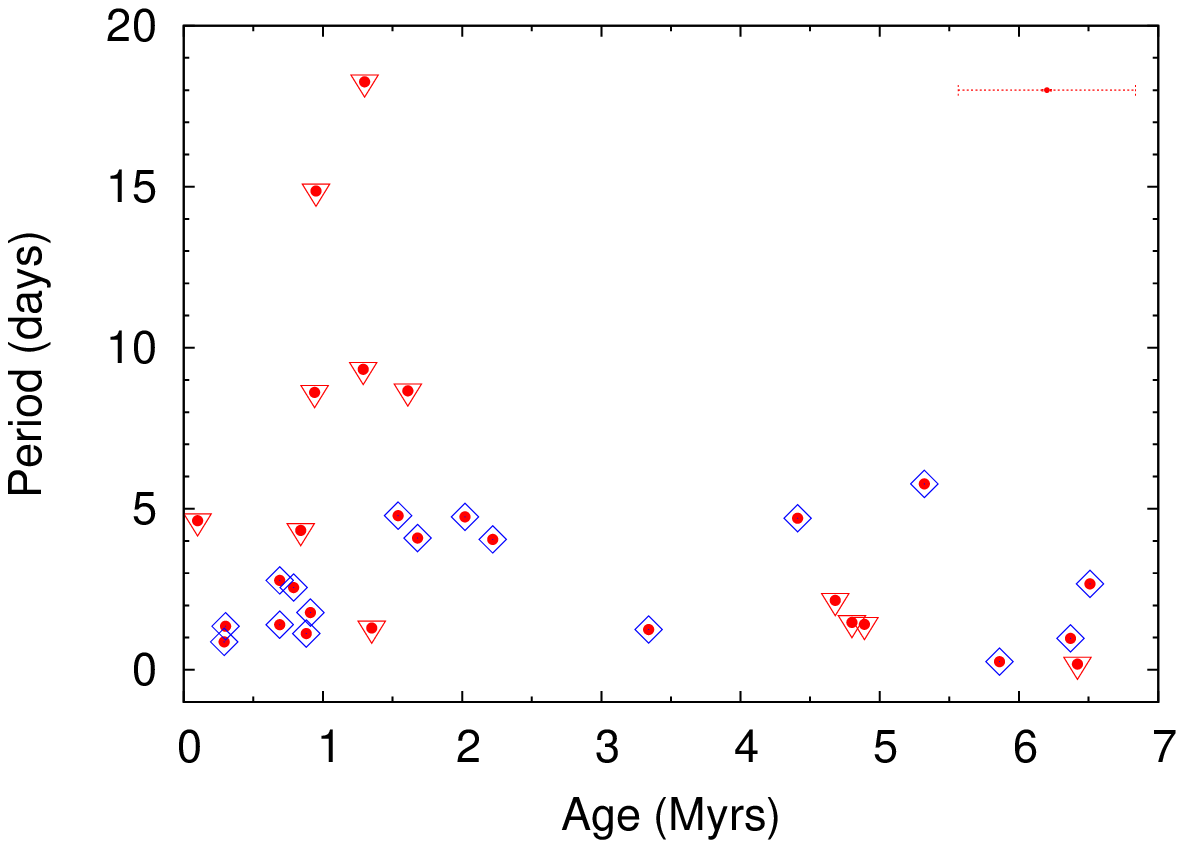}
\hspace{0.5 cm}
\includegraphics[width=0.37\textwidth]{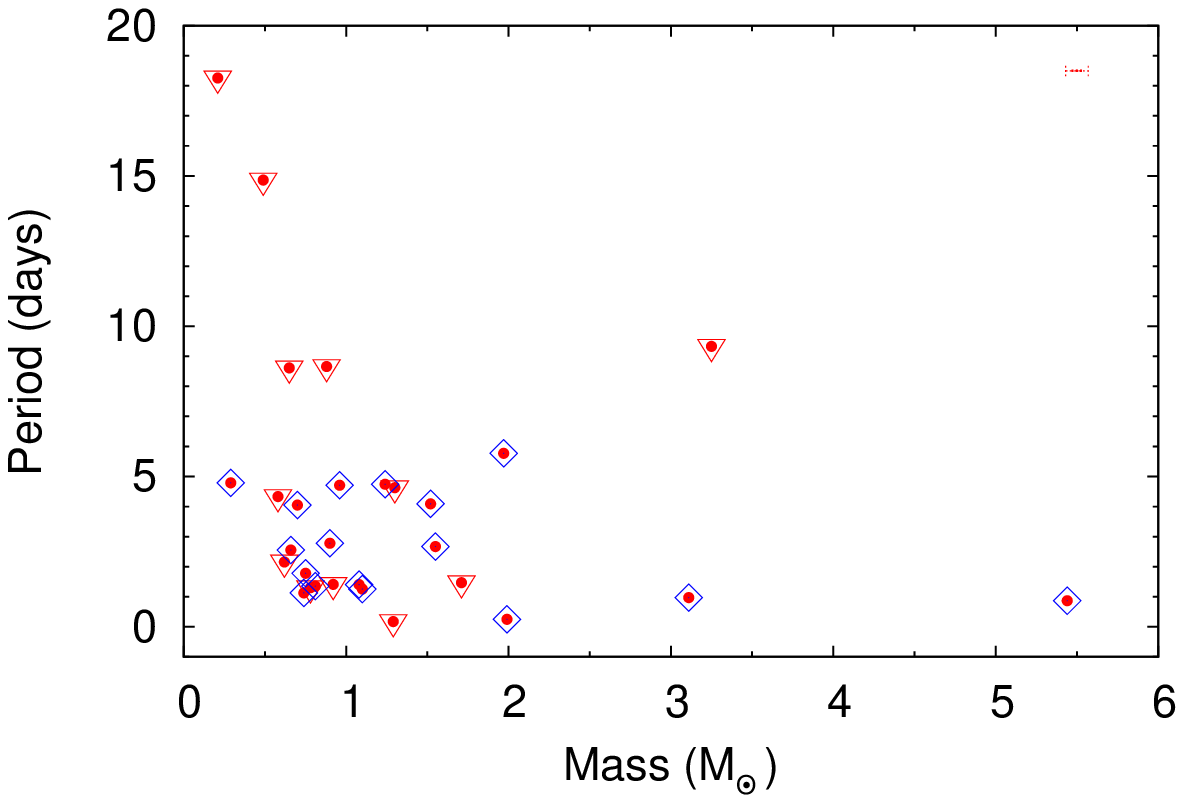}
 \caption{Rotation period as a function of age and mass of PMS variable stars. The Class\,{\sc ii} and Class\,{\sc iii} sources 
are represented with inverted triangles and diamonds, respectively. Corresponding mean errors are shown with error bars in the 
top right corner of each panel.}
 \label{fig : Period_Age-Mass}
\end{figure*}

\begin{figure*}
\centering
\includegraphics[width=0.35\textwidth]{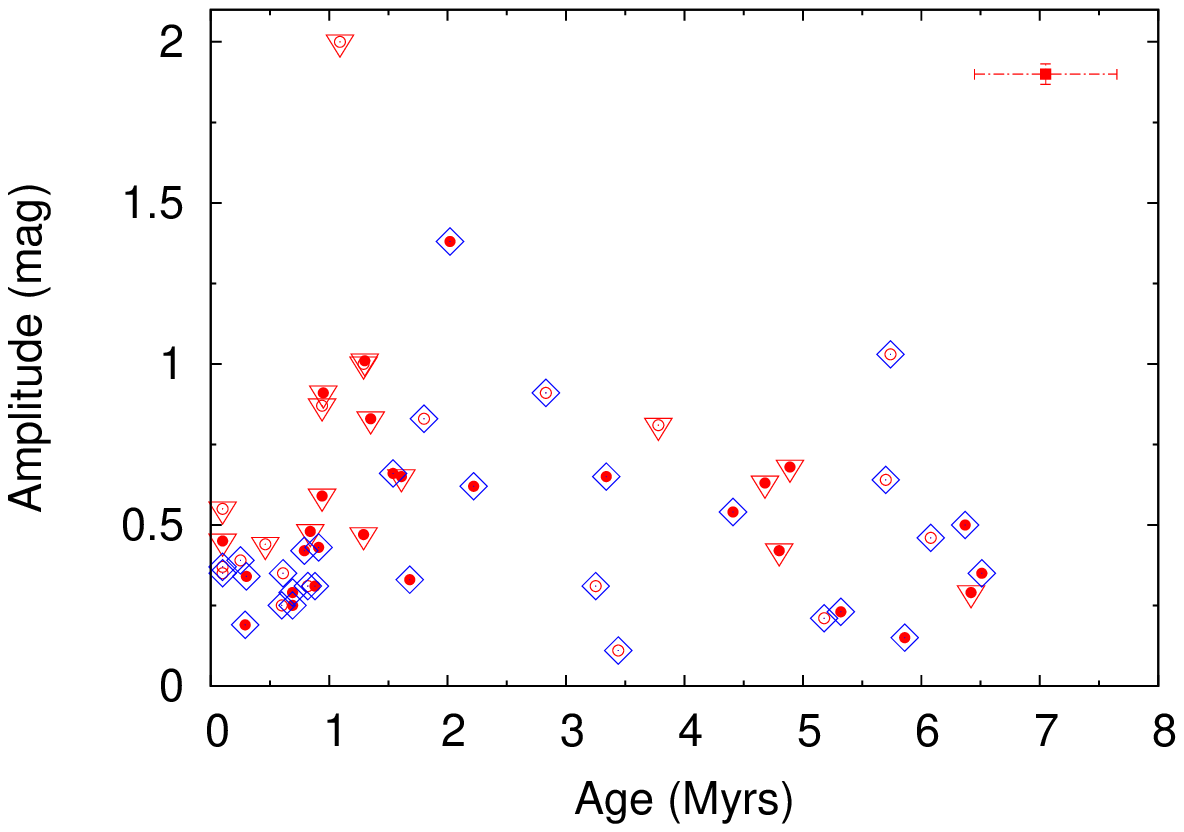}
\hspace{0.5 cm}
\includegraphics[width=0.37\textwidth]{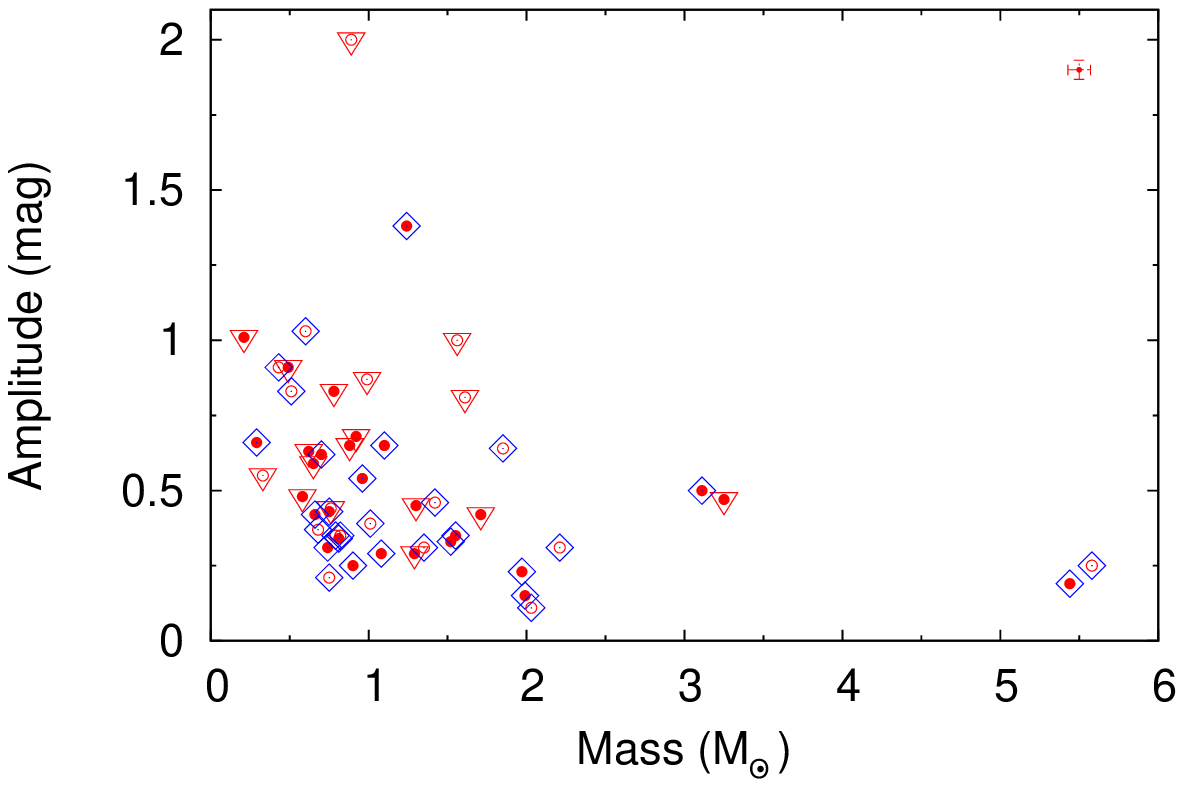}
 \caption{Amplitude of variations as a function of age and mass of PMS variable stars. Filled and open circles represent periodic 
and non-periodic variables, respectively. Other symbols are the same as in Fig. \ref{fig : Period_Age-Mass}.  Corresponding mean 
errors are shown with error bars.}
 \label{fig : Amplitude_Age-Mass}
\end{figure*}

\subsection{Correlation of periods/amplitudes of variables with their stellar parameters}
\label{age_mass_correlation}

To check the dependence of the variation of the PMS variables on
their evolutionary status, in  Fig. \ref{fig: CAD} we plot the
normalized cumulative amplitude  distribution of Class\,{\sc ii} and Class\,{\sc iii} sources.
This distribution clearly indicates that Class\,{\sc ii} sources exhibit larger 
amplitude variations as compared to Class\,{\sc iii} sources 
with a  96\% confidence level as calculated by the Kolmogorov–Smirnov test. Similar results have been reported 
by \citet{2011MNRAS.418.1346L,2012MNRAS.427.1449L} and  \citet{2019arXiv190600256B}. 
To further investigate how the period of variability evolves with the age and how the mass  of PMS stars 
influence their periods, we  plot the period of the PMS variables as a function of age and mass 
in the left and right panels of Fig. \ref{fig : Period_Age-Mass}, respectively. 
The left panel of Fig. \ref{fig : Period_Age-Mass} indicates that stars with periods up to $\sim$6 days 
are uniformly distributed for the entire range of ages of the PMS sources, 
whereas all the PMS variables  having periods $>$ 6 days are younger than 2 Myr.   
\citet{2014MNRAS.442..273L,2016MNRAS.456.2505L} have also found a similar result that PMS stars 
with age $\geqslant$ 3 Myr seem  to be relatively fast rotators. 
The right panel of Fig. \ref{fig : Period_Age-Mass} displays the dependence of period on the stellar mass. 
Although the sample is small, but we can see a trend in the distribution: 
out of 5 slow rotators (period $>$ 6 days), 4 have masses less than 1 M$_\odot$. 
and variables with periods $<$ 6 days have masses $\leqslant$ 2 M$_\odot$, 
whereas out of 3 high mass variables (M $>$ 3 M$_\odot$), 2 
have periods $\sim$1 days. A similar trend has been reported by 
\citet{2014MNRAS.442..273L,2016MNRAS.456.2505L}. \citet{2005A&A...437..637L} have also found a  
strong correlation between stellar mass and rotation rate in the case of IC 348. 
They concluded that the strong mass dependence of rotation 
rate seen in ONC \citep{2000AJ....120..349H} may well be a common feature of young stellar populations. 

Fig. \ref{fig : Amplitude_Age-Mass} plots amplitude of variability as a function of age and 
mass of the PMS variables,  which roughly indicates that their  amplitudes  
decrease with the increase in mass and age. This is similar to the previous findings  
\citep{2000AJ....120..349H,2002A&A...396..513H,2005A&A...437..637L,2014MNRAS.442..273L,2016MNRAS.456.2505L}.  
The amplitude decrease with age could be due to the dispersal of the disc. The present result further supports that of our 
previous studies \citep{2011MNRAS.418.1346L,2012MNRAS.427.1449L,2016MNRAS.456.2505L} that the disc dispersal 
mechanism is less efficient for relatively low mass stars and that a significant amount of discs is dispersed by $\sim$ 5 Myr. 
This is also in accordance with the result obtained by \citet{2001ApJ...553L.153H}.

\begin{figure*}
\centering
\includegraphics[width=0.35\textwidth]{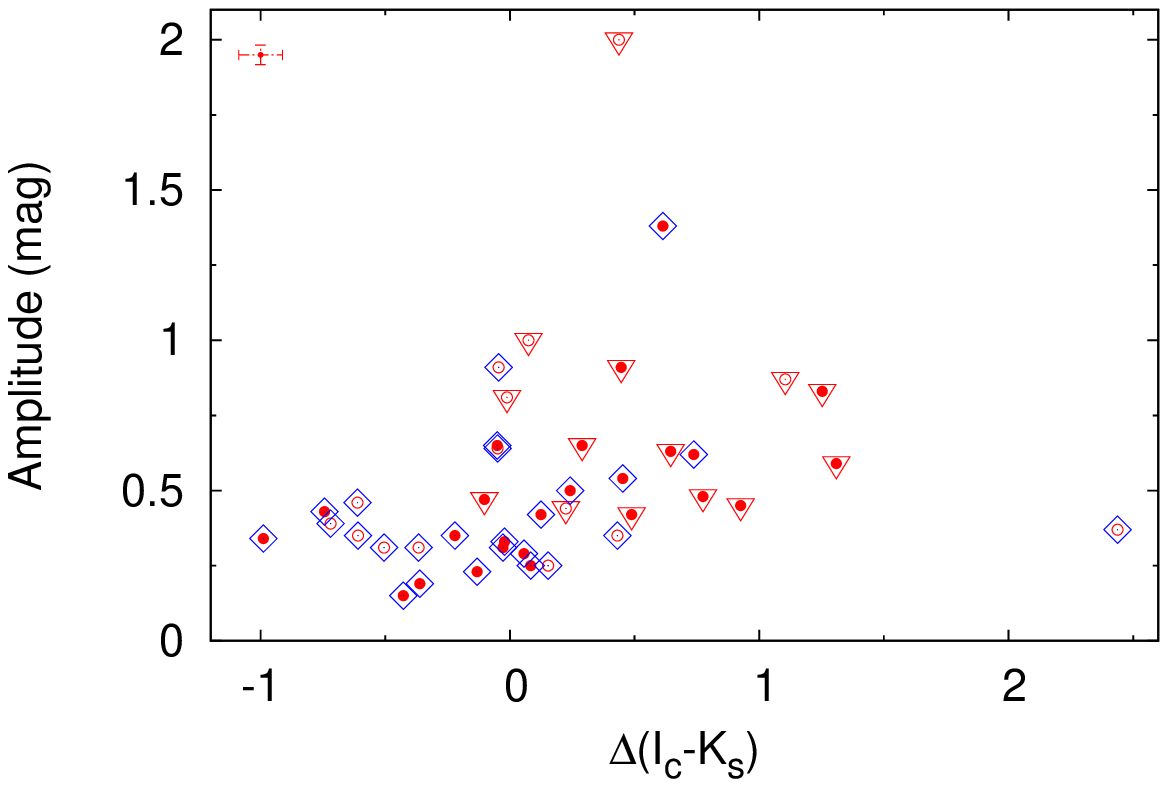}
\hspace{0.5cm }
\includegraphics[width=0.335\textwidth]{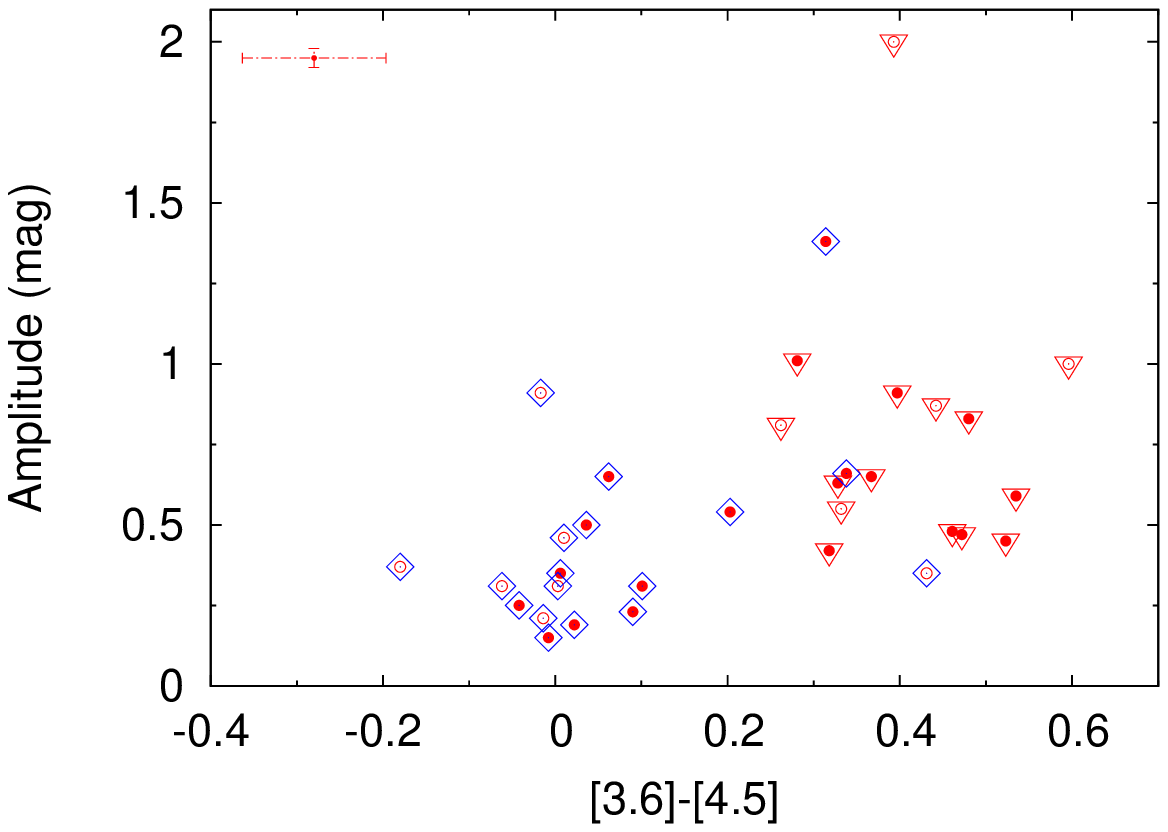} \\
\vspace{0.2 cm}
\hspace{0.2cm }
 \includegraphics[width=0.35\textwidth]{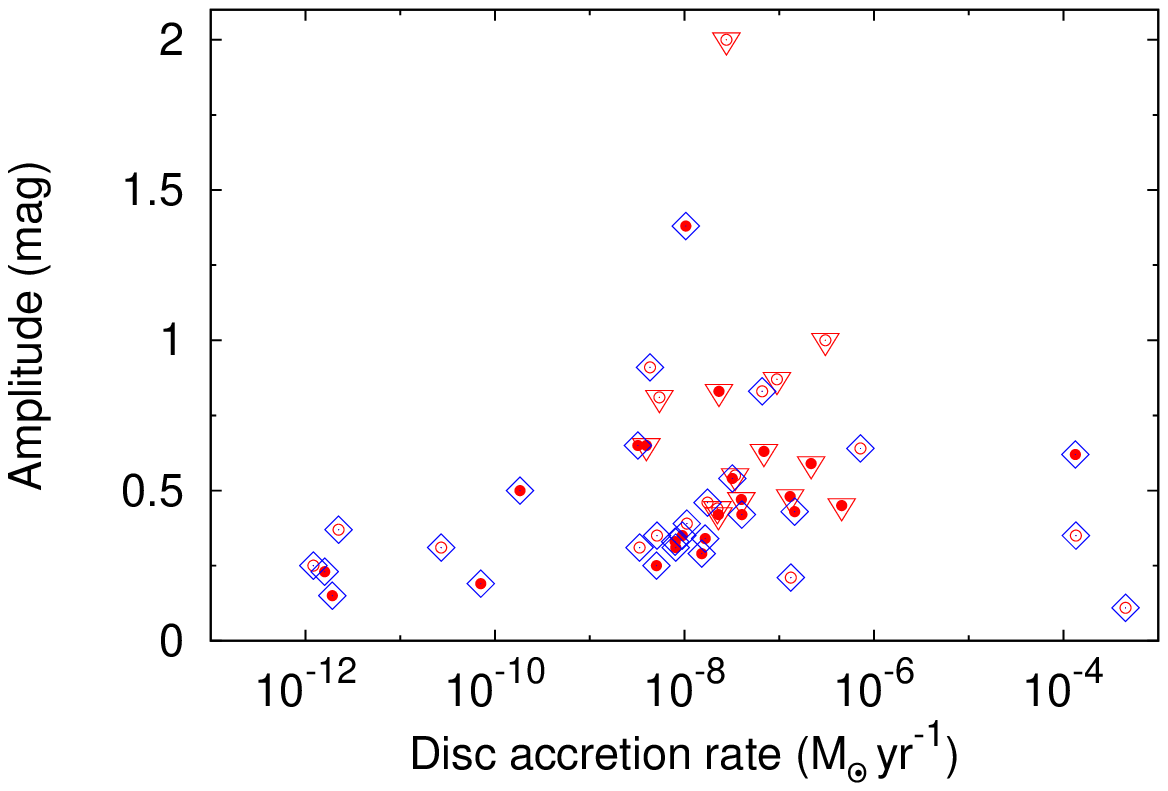}
 \hspace{0.5cm }
 \includegraphics[width=0.355\textwidth]{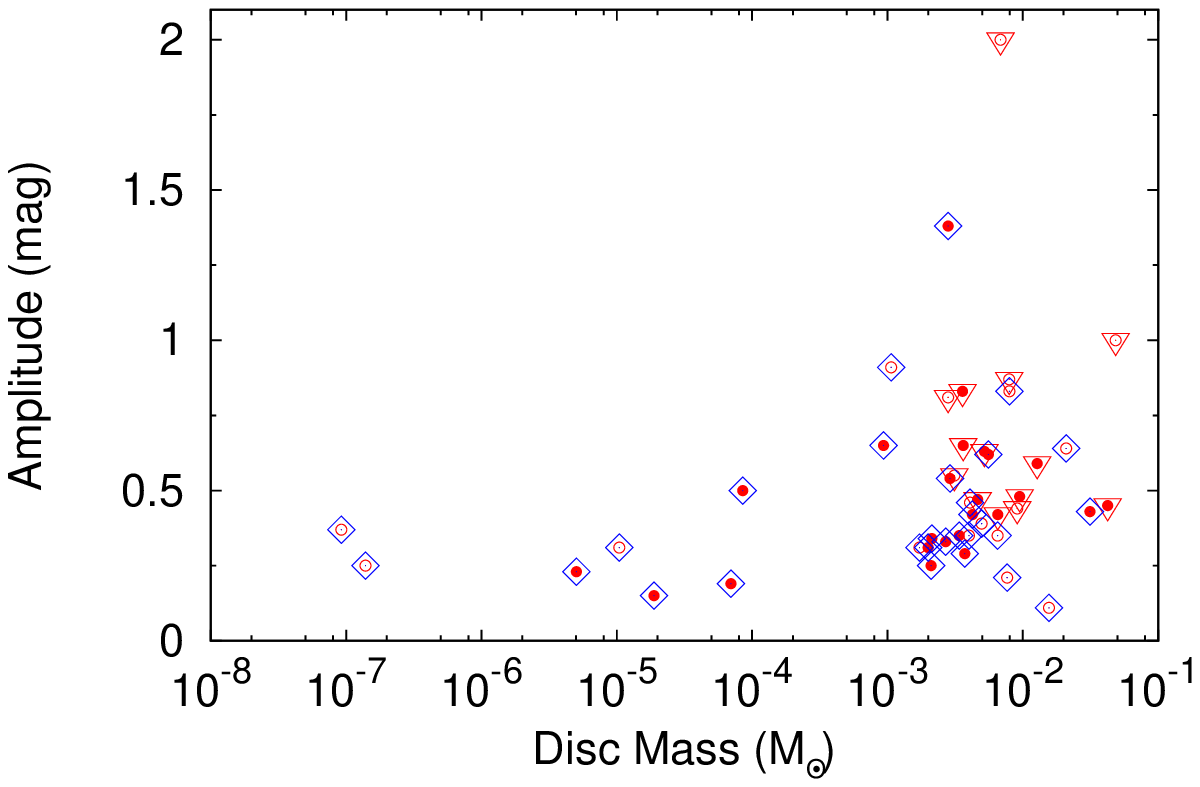}
 \caption{Top panels : Variation amplitude of the PMS variables as a function of NIR excess $\Delta (I_{c}-K_{s})$ and MIR 
excess $[[3.6]-[4.5]]$. Bottom panels : Variation amplitude as a function of disc accretion rate and disc mass. The symbols are 
the same as in Fig. \ref{fig : Amplitude_Age-Mass}.}
  
 \label{fig : Amplitude_NIR-MIR-Accretion-Discmass}
\end{figure*}

\begin{figure*}
\centering
   \includegraphics[width=0.35\textwidth]{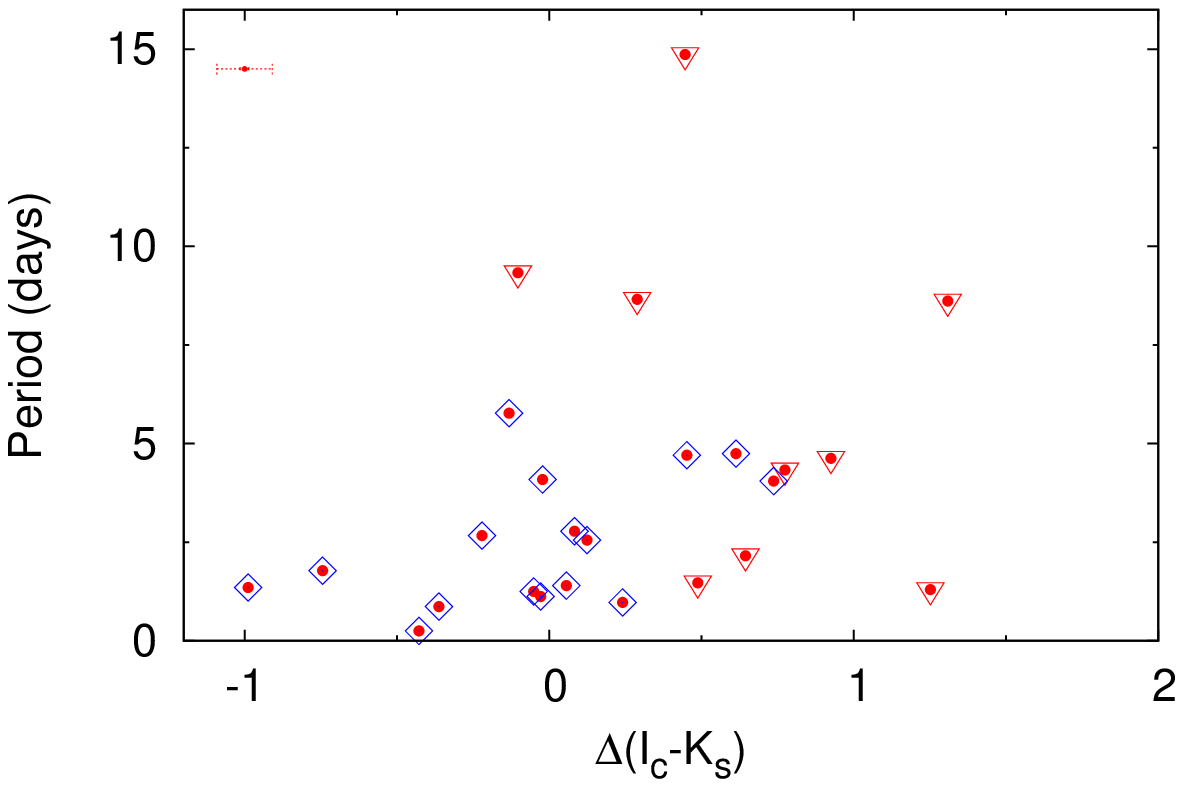}
   \hspace{0.4 cm}
  \includegraphics[width=0.33\textwidth]{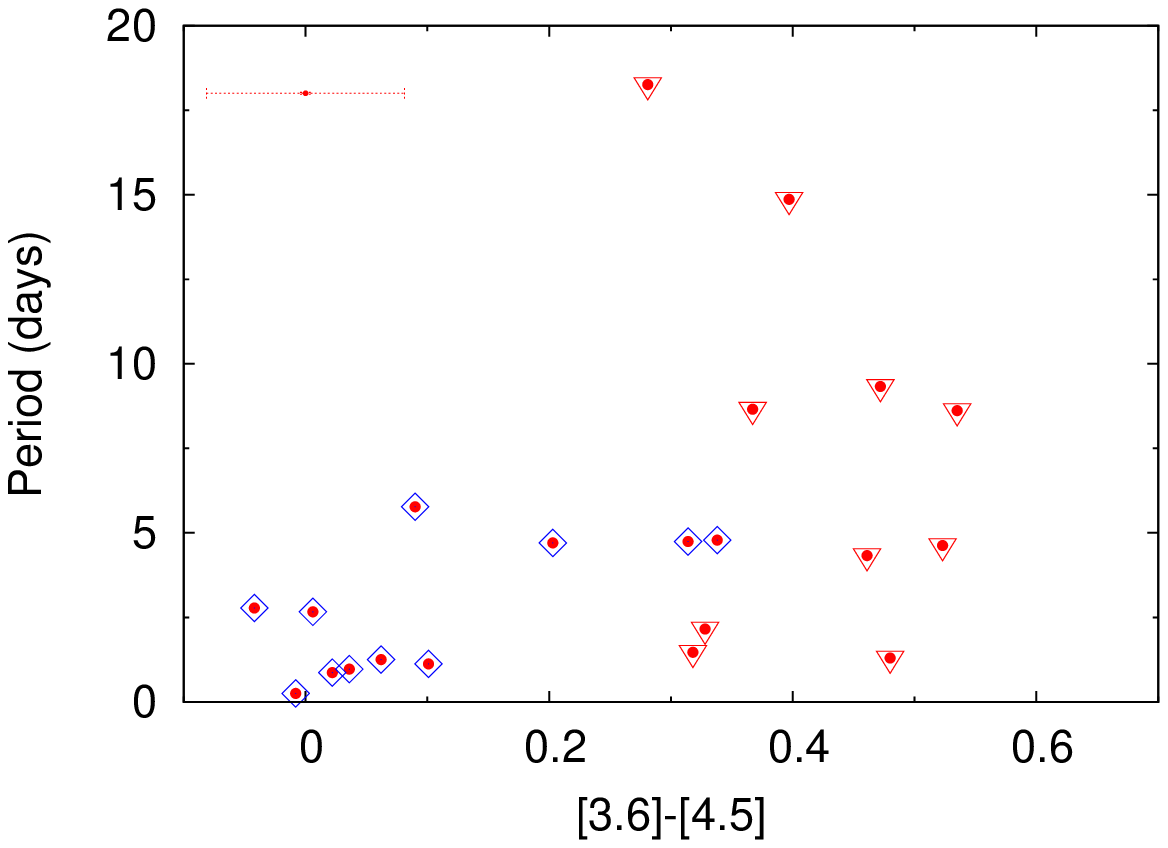} 
  \vspace{0.05 cm}
   \includegraphics[width=0.33\textwidth]{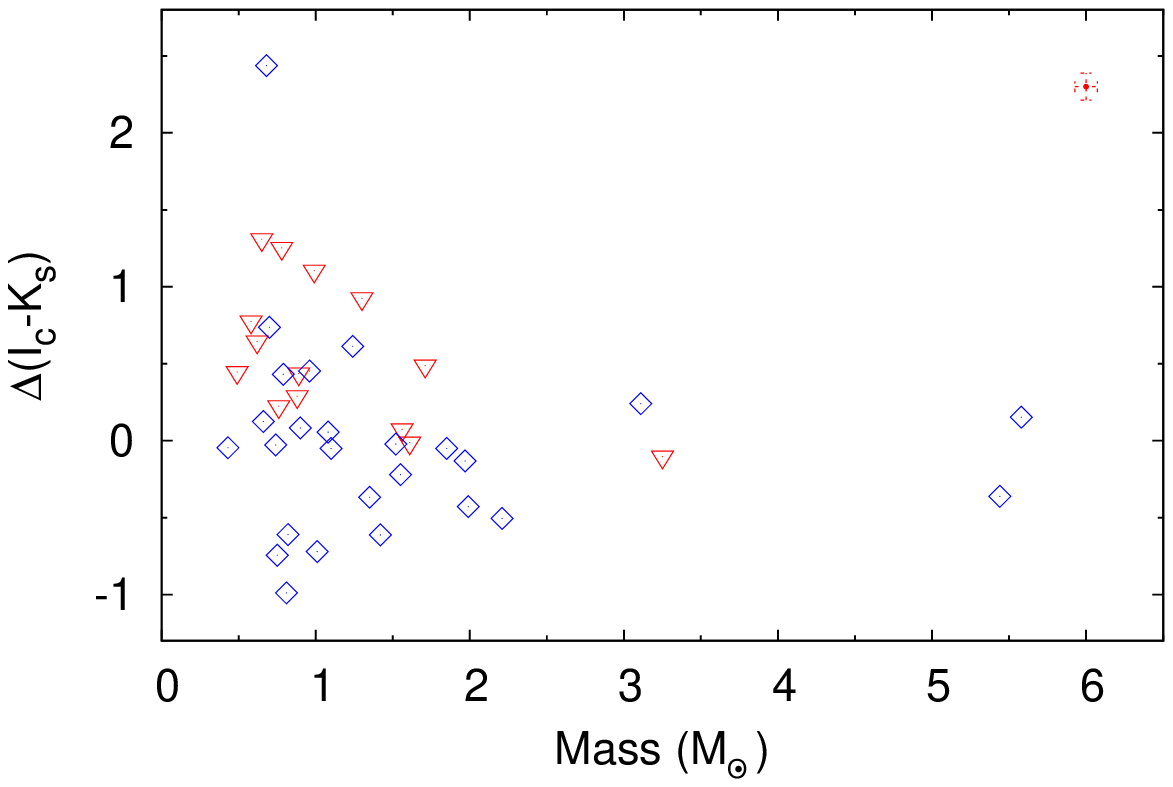}
 \hspace{0.5 cm}
 \includegraphics[width=0.33\textwidth]{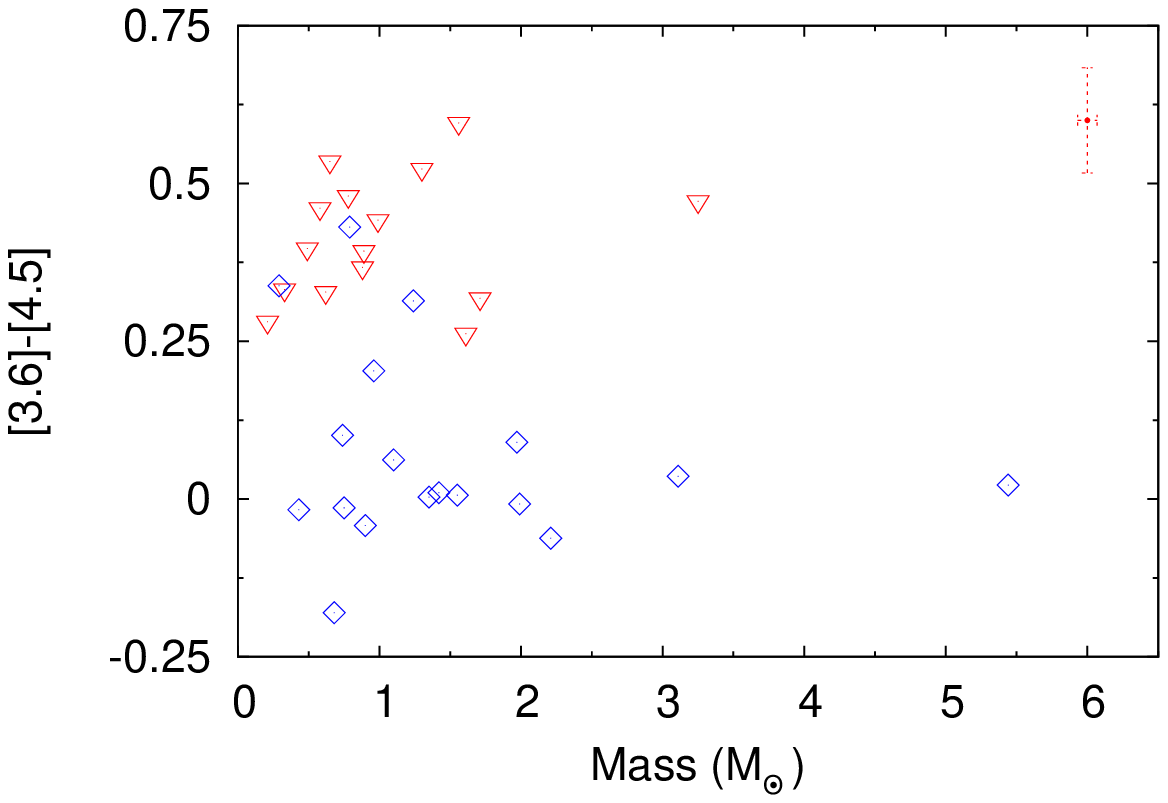}

  \caption{Top panels : Rotation period of PMS variables as a function of NIR excess $\Delta (I_{c}-K_{s})$ and MIR excess $[[3.6]-[4.5]]$. 
Bottom Panels: NIR excess $\Delta (I_{c}-K_{s})$ and MIR excess $[[3.6]-[4.5]]$ as a function of stellar mass. Corresponding mean errors are 
shown with error bars. The symbols are the same as in Fig. \ref{fig : Amplitude_Age-Mass}.}
 
   \label{fig : Period_NIR-MIR_&_NIR-MIR_Mass}
\end{figure*}

\subsection{Correlation of periods/amplitudes of PMS variables with disc evolution}

To understand the influence of circumstellar discs on the period and amplitude of young stars,
we have to first identify a suitable disc indicator. Various disc indicators, such as 
equivalent  widths of $H\alpha$ emission line  and  Ca II triplet lines, 
$\Delta (H-K_{s})$ and $\Delta (I_{c}-K_{s})$ indices, disc fraction, etc.,
have been used in the previous studies  \citep[e.g.,][]{2000AJ....120..349H,2005A&A...437..637L,2007ApJ...671..605C}. 
Since $I_{c}$ band fluxes originate solely from the photosphere unlike $K_{s}$~band flux which originates in 
circumstellar disc emission as well as photospheric emission, 
this provides $(I_{c}-K_{s})$ index with a longer wavelength base compared to other NIR indices
(e.g. $(J-H)$ or $(H-K_{s})$). This index is expressed as \citep[cf.][]{1998AJ....116.1816H}:

\begin{equation}
\Delta(I_{c}-K_{s}) = (I_{c}-K_{s})_{obs} - (A_{I_{c}} - A_{K_{s}}) - (I_{c}-K_{s})_{0} 
\end{equation}

where $(I_{c}-K_{s})_{obs}$ is the observed colour of the star, $(I_{c}-K_{s})_{0}$ is the intrinsic colour  and $A_{I_{c}}$ and 
$A_{K_{s}}$ are the interstellar extinction in the $I_{c}$ and $K_{s}$ bands, respectively. 
We have used the  $A_{V}$ value ($2.6\pm0.3$ mag) as determined by using $Q$ parameter \citep{1953ApJ...117..313J}
to correct the stars for their extinction.
Intrinsic colour $(I-K)_{0}$ of the variables were taken from the PMS isochrones of \citet{2000A&A...358..593S}  according to 
the age and mass as derived by their CMD and then corrected for the distance and extinction.

Since $\Delta(I_{c}-K_{s})$  is sensitive to the inner part of the disc, 
we also used the available MIR data of these variables to compute an 
MIR index $[[3.6]-[4.5]]$, which is a better indicator of the presence of 
disc and its evolution (\citealt{2000AJ....120.3162L}; \citealt{2010A&A...515A..13R}). Since the presence of disc also induces  
the accretion activity, we have also used the information such as the disc accretion rate 
and the disc mass obtained from the SED fitting tool  (cf. Sec \ref{SED}). 
The upper panels of Fig. \ref{fig : Amplitude_NIR-MIR-Accretion-Discmass} show the variation of amplitude 
as a function  of $\Delta (I_{c}-K_{s})$ and $[[3.6]-[4.5]]$. 
A trend can be clearly seen in the sense that the larger  values of the disc indicators, i.e., $\Delta (I_{c}-K_{s})$ or $[[3.6]-[4.5]]$  
correspond to relatively larger amplitude variations.  The $[[3.6]-[4.5]]$ versus amplitude diagram manifests that Class\,{\sc ii} 
sources have  active circumstellar discs  as compared to Class\,{\sc iii} sources, hence  these sources  
display more active and dynamic variability due to the accretion activities. 
In the case of  NGC 2282, \citet{2018MNRAS.476.2813D} have also reported  a positive correlation between RMS amplitudes and 
$\Delta(I_{c}-K_{s})$/$[[3.6]-[4.5]]$ disc indicators. The lower panels of Fig. \ref{fig : Amplitude_NIR-MIR-Accretion-Discmass} 
show amplitudes as a function of the disc accretion rate and disc mass, which indicates that 
higher disc accretion activity ($>10^{-8}$ M$_\odot~ yr^{-1}$) induces higher 
amplitude variation. The amplitude variation also depends on the mass of the stellar 
disc in the sense that discs with mass $\geqslant$ $10^{-3}$ M$_{\odot}$  show relatively larger amplitude variation.

\begin{figure*}
\centering
 \includegraphics[width=0.35\textwidth]{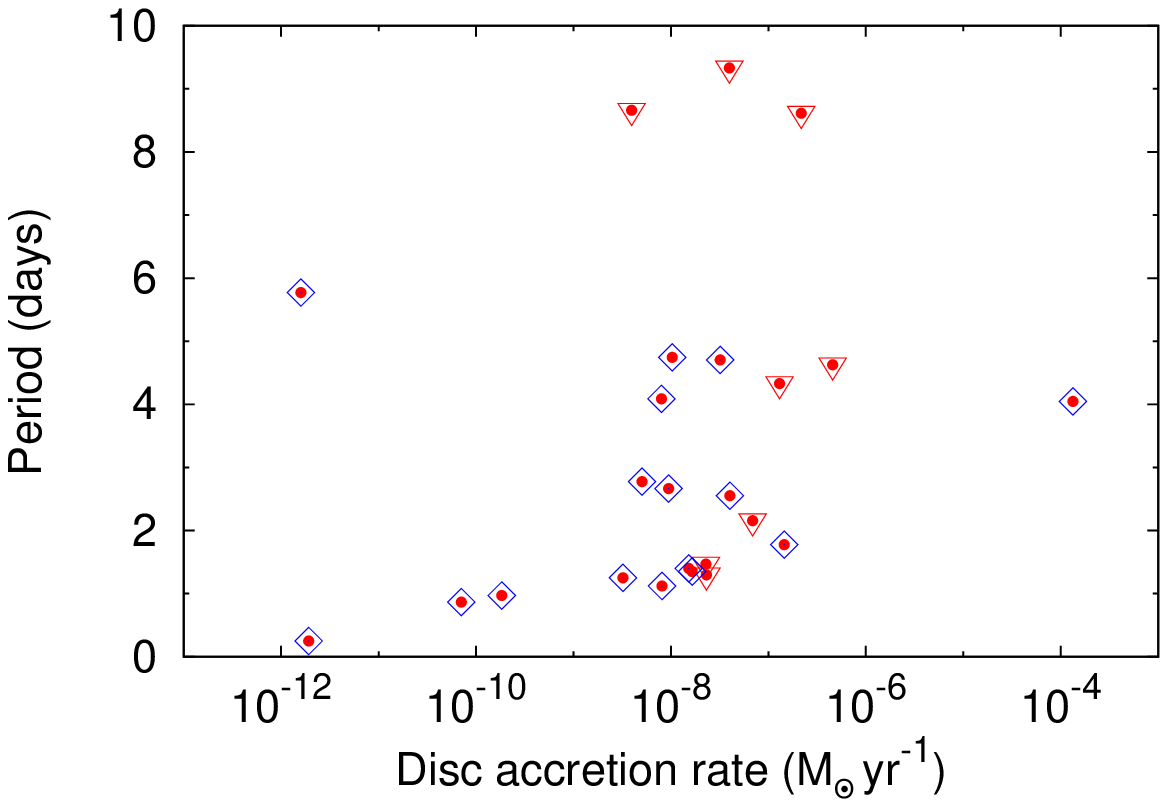}
 \hspace{0.4 cm}
 \includegraphics[width=0.365\textwidth]{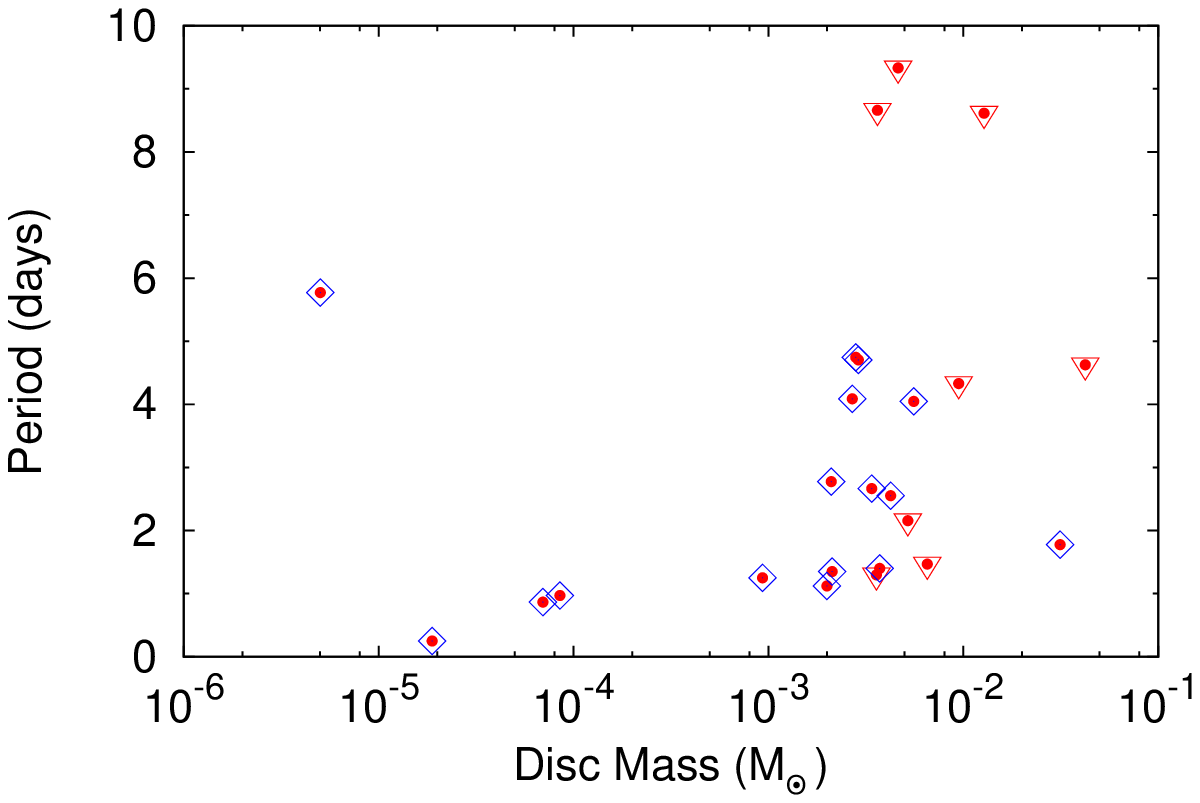}
 \caption{ Rotation period as a function of disc accretion rate and disc mass. The symbols are the same as 
in Fig. \ref{fig : Amplitude_Age-Mass} }
 \label{fig : Period_Accretion-Discmass}
\end{figure*}

The extensive study of rotation rates in ONC carried out  
by \citet{2000AJ....120..349H,2002A&A...396..513H} reveals a strong correlation between 
rotation period and infrared excess, suggesting that the observed rotation 
period distribution could be due to the disc-locking mechanism. 
In Fig. \ref{fig : Period_NIR-MIR_&_NIR-MIR_Mass} (upper panels) we plot rotation period as a 
function of the $\Delta (I_{c}-K_{s})$ and $[[3.6]-[4.5]]$ disc indicators. 
The $[[3.6]-[4.5]]$ MIR disc indicator clearly suggests that all the five 
slow rotators (having periods $>$ 6 days) are Class\,{\sc ii} sources  having 
higher value of $[[3.6]-[4.5]]$ index ($\sim$0.4 mag), 
whereas Class\,{\sc iii} sources with $[[3.6]-[4.5]]$  $\sim$0.0 mag have periods $\leqslant$ 6 days.   
This result indicates a reasonable likelihood that the presence of disc affects 
the rotation of the star. Similar results were reported by \citet{1993AJ....106..372E} 
and the physical interpretation proposed by  the authors was that the discs 
slows the rotation of stars through magnetic interaction \citep{1991ApJ...370L..39K,1995ApJ...447..813O}.
Hence, stars having discs either locked or recently 
released tends to be slower rotators than stars which already dispersed their disc. 
However, \citet{2005A&A...437..637L} did not find any correlation between the 
$H\alpha$ equivalent width or $(K_{s}-L)$ excess and the rotation period of the stars. 
They cautioned that  the correlation between rotation period and infrared excess
expressed by $(I_{c}-K_{s})$ index can also arise due to the strong dependence of $(I_{c}-K_{s})$ 
on stellar mass in  the sense that high-mass stars have larger $(I_{c}-K_{s})$ values as compared to 
low-mass stars \citep{1998AJ....116.1816H}. 
To check the dependence of  $\Delta (I_{c}-K_{s})$ and $[[3.6]-[4.5]]$ NIR/ MIR excess  on stellar mass, we plot these parameters 
in the bottom panels of Fig. \ref{fig : Period_NIR-MIR_&_NIR-MIR_Mass}. The plots indicate a weak correlation between 
NIR/ MIR excess with stellar mass in the sense that higher values of NIR/MIR excess 
are associated with relatively low mass stars. Hence relatively low mass stars are preferably slow rotators.
 
\citet{2007ApJ...671..605C} reported a clear increase of the disc fraction with period of stars
in ONC and NGC 2264. They showed that the long-period peak (P $\sim$ 8 days) 
of the bimodal distribution observed in ONC is dominated by the population of stars 
possessing a disc, while the short-period peak (P $\sim$ 2 days) is dominated by the discless
population and concluded that this is a strong evidence that the star-disc interaction regulates the angular momentum 
of young stars. 

 Fig. \ref{fig : Period_Accretion-Discmass}  shows the effect of disc accretion rate and stellar disc mass on 
the period. It is evident from this figure that both the disc accretion rate and stellar 
disc mass affects the stellar rotation in the sense that slow rotators are highly accreting stars 
(disc accretion rate $\gtrsim$  5$\times 10^{-9}$ M$_\odot ~ yr^{-1}$ ) and have higher stellar disc mass  ($\gtrsim$ 3 $\times 
10^{-4}$ M$_\odot$ ).

\section {Summary and conclusion}
\label{sect:conclusion}

In this paper we have presented the multi-epoch deep $I_{c}$ band ($\sim$20 mag) photometric monitoring of the Sh 2-170 region to 
understand the characteristics of variables in the region. Following are the main results.

\begin{itemize}
\item
We have identified 71 variables in the region. The probable members associated with Sh 2-170 are identified on the basis of their 
PM, location in the optical CMD and presence or absence of excess IR emission. 
Forty nine variables are found to be probable PMS stars, whereas remaining 17 stars are found to be MS/field population. 
Of 49 PMS variables, 17 and 32 are classified as Class\,{\sc ii} and Class\,{\sc iii} sources, respectively. Ten and nineteen of 
Class\,{\sc ii} and Class\,{\sc iii} variables are found to be periodic; 15 MS variables are found to be periodic.

\item
The majority of the PMS variables have mass and age in the range of 0.2 $\leq$ M/M$_\odot$ $\leq$ 3.0 and  0.1 - 2.0 Myrs, 
respectively, and hence should be TTSs. The rotation period of the  PMS variables  ranges from 4 hrs  to 18 days whereas 
the amplitude varies  from 0.1 mag to 2.0 mag. The amplitude is larger in  Class\,{\sc ii} sources (up to 2.0 mag) as compared 
to those in Class\,{\sc iii} sources ($\sim$0.7 mag). 

\item
The period distribution of the PMS stars reveals a bimodal nature with peaks at $\sim$1.5 days and 4.5 days.  
The period for Class\,{\sc ii} sources ranges  from 1 to 18 days, whereas Class\,{\sc iii} sources have periods in the range 
4 hrs - 6 days.  The correlation between stellar rotation period and NIR/MIR excess is apparent in the sense that slow rotators 
have larger  NIR/MIR excesses as compared to fast rotators. It is also found that the disc accretion rate and stellar disc mass 
are correlated to the stellar rotation in the sense that highly accreting stars (disc accretion rate $\gtrsim$ 5$\times 
10^{-9}$ M$_\odot yr^{-1}$) as well as stars of higher disc mass ($\gtrsim$ 3 $\times 10^{-4}$ M$_\odot$) are preferably slow 
rotators.

\item
The bimodal period distribution and dependence of rotation period on IR excess/accretion rate/disc mass  
are compatible with the disc-locking model. This model suggests that when a star is disc-locked, its rotation speed doesn't 
change and  when the star is released from the locked-up disc, it can spin up with its contraction.

\item
Both the period and amplitude of PMS variables decrease with increasing stellar mass. This is in accordance with our earlier 
proposition that the decrease in variability amplitude in relatively massive stars could be due to the dispersal of circumstellar disc
and that the mechanism of this disc dispersal operates less efficiently in relatively low mass 
stars \citep{2011MNRAS.418.1346L,2014MNRAS.442..273L,2016MNRAS.456.2505L}.

\item
The amplitude of variables is correlated with the NIR and MIR disc indicators in the sense that the larger value of disc 
indicators, i.e., $\Delta(I_{c}-K_{s})$ or $[[3.6]-[4.5]]$, corresponds to relatively larger amplitude variation. The $[[3.6]-[4.5]]$  disc 
indicator manifests that Class\,{\sc ii} sources have active circumstellar disc around them as compared to Class\,{\sc iii} sources, 
hence these sources are more active and dynamic and display larger variability due to the accretion activities. The amplitude of 
variability also depends on the disc accretion rate and disc mass in the sense that higher disc accretion activity 
($\gtrsim 10^{-8}$ M$_\odot yr{^{-1}}$) and massive discs of $\gtrsim 10^{-3}$ M$_\odot$ induce higher amplitude variation.

\item
These results are also in agreement with the notion that cool spots on WTTSs are responsible for most or all of their variations, while 
hot spots on CTTSs resulting from variable mass accretion from the inner disc contribute to their larger amplitudes and more 
irregular behaviours. 
\end{itemize}

Here, we would like to emphasize that 
cool spots on the surface of a WTTS can change their shapes and distribution on a timescale of months, 
which makes it impossible to maintain the integrity of phase over different observing seasons \citep{2000AJ....120..349H}. 
In future studies, observations in a larger  number of nights preferably within an observing season will be useful 
to determine the periods of PMS stars more accurately and to obtain statistically more significant 
samples of PMS variables. Also desirable are the observing in different cluster environment.

\section*{Acknowledgment}
 We are very thankful to the referee Prof. William Herbst for the critical review of the manuscript and useful comments.
The observations reported in this paper were obtained by using the
 1.3m Devasthal Fast Optical Telescope (DFOT, India), 0.7m TRT-GAO (Thai Robotic Telescope at Gao Mei Gu Observatory, China)
and 0.5m telescope of Thai National Observatory (TNO, Thailand).
This work made use of data
from the Two Micron All Sky Survey (a joint project of the University of Massachusetts and the Infrared Processing and Analysis
Center/California Institute of Technology, funded by the National Aeronautics and Space Administration and the National Science
Foundation), and archival data obtained with the $Spitzer$ Space Telescope and Wide Infrared Survey Explorer (operated by 
the Jet Propulsion Laboratory, California Institute of Technology, under contract with the NASA.
This publication also made use of data from the European Space Agency (ESA) mission $Gaia$
(https://www.cosmos.esa.int/gaia), processed by the $Gaia$ Data Processing and Analysis Consortium (DPAC,
https://www.cosmos.esa.int/web/gaia/dpac/consortium). Funding for the DPAC has been provided by national institutions, 
in particular the institutions participating in the $Gaia$ multilateral agreement. AKP is thankful to NARIT, Thailand for the support 
during the stay at NARIT.

\bibliography{stock}{}
\bibliographystyle{mnras}

\clearpage
\onecolumn
\begin{landscape}
\begin{table*}
\centering
\caption{ Sample of stars identified as members of the Stock 18 cluster.
The complete table is available in the electronic form only.\label{PMT} }
\begin{tabular}{@{ }c@{ }c@{ }r@{ }c@{ }c@{ }c@{ }c@{ }c@{ }c@{ }c@{ }c@{ }c@{ }}
\hline
ID& $\alpha_{(2000)}$&$\delta_{(2000)}$&$V$ & $I_{c}$ & Flag&$\mu_\alpha\pm\sigma$&$\mu_\delta\pm\sigma$ &  Parallax$\pm\sigma~~~~~$& Probability& $G$ & $G_{BP}-G_{RP}$ \\
& {\rm $(degrees)$} & {\rm $(degrees) $} &  (mag) & (mag)& &  (mas/yr)& (mas/yr)& (mas) & (\%) & (mag) & (mag) \\
\hline
    M1  &   0.394085  &   64.623123  &    ---            ---    &    ---            ---    &  ---     &  -2.6403 $\pm$   0.0538  & -0.5290 $\pm$   0.1046  &  0.2530  $\pm$  0.0400  &  100  &  12.525  &   ---  \\ 
    M2  &   0.407023  &   64.626556  &   18.430  $\pm$   0.071  &   16.258  $\pm$   0.043  &    a     &  -2.6724 $\pm$   0.1730  & -0.4744 $\pm$   0.1477  &  0.3230  $\pm$  0.1050  &   99  &  17.800  &  1.921 \\
    M3  &   0.446222  &   64.629654  &    ---            ---    &    ---            ---    &  ---     &  -3.1729 $\pm$   0.9959  & -0.4800 $\pm$   0.6734  & -0.0480  $\pm$  0.4665  &   85  &  20.017  &  2.091 \\
    M4  &   0.441568  &   64.624008  &  19.676   $\pm$   0.054  &   17.184  $\pm$   0.034  &    a     &  -3.0820 $\pm$   0.4108  & -0.5451 $\pm$   0.2632  &  0.1190  $\pm$  0.1898  &   95  &  18.693  &  2.219 \\
\hline
\end{tabular}
\\
\hspace{- 4.9 cm}  a : $V$ and $I_{c}$ data taken from \citet{2012NewA...17..160B}. ;
 b : $V$ and $I_{c}$ data taken from the present observations.\\

\end{table*}

\begin{scriptsize}
\centering
\begingroup
\renewcommand\arraystretch{1.2}
\setlength{\tabcolsep}{1.5pt}
\begin{longtable}[c]{ccccccccccccccccccc}
\caption{Sample of identified variables and their physical parameters as derived from CMD and/or SED. The complete table 
is available in the electronic form only. \label{Variables}} \\

 \multicolumn{19}{}{Table}\\
 \hline

 ID &  $\alpha_{(2000)}$ & $\delta_{(2000)}$ &  $V$   &  $I_{c}$   &  Flag &  Period &  Amplitude &  RMS &  $\Delta(I_{c}-K_{s})$ &  $[3.6]-[4.5]$ & Age$\dagger$ & Mass$\dagger$     &  Data points* &  $\chi^{2}*$ &  Age* &  Mass*    &  Disc mass* &  Disc accretion rate* \\
{}   &  (degrees) &  (degrees) &  (mag) &  (mag) & {}     &  (days)   & (mag) &  (mag) & (mag) &  (mag)           &  (Myr) &  (M$_\odot$) & {}            & {}       &   (Myr)  &  (M$_\odot$) &  (M$_\odot$) & (M$_\odot$~$yr^{-1}$) \\

 \endfirsthead
 
 \hline
 \multicolumn{19}{c}{Continuation of Table \ref{Variables}}\\
 \hline 
 
 ID &  $\alpha_{(2000)}$ & $\delta_{(2000)}$ &  $V$   &  $I_{c}$   &  Flag &  Period &  Amplitude &  RMS &  $\Delta(I_{c}-K_{s})$ &  $[3.6]-[4.5]$ & Age$\dagger$ & Mass$\dagger$     &  Data points* &  $\chi^{2}$* &  Age* &  Mass*    &  Disc mass* &  Disc accretion rate* \\
{}   &  (degrees) &  (degrees) &  (mag) &  (mag) & {}     &  (days)   & (mag) &  (mag) & (mag) &  (mag)           &  (Myr) &  (M$_\odot$) & {}            & {}       &   (Myr)  &  (M$_\odot$) &  (M$_\odot$) & (M$_\odot$~$yr^{-1}$) \\

\hline
 \endhead
 
 \hline
 \endfoot
 
 \hline
 \multicolumn{19}{}{}\\

 \endlastfoot
\hline

       V1  &   0.343856   &  64.637985  &  13.917$\pm$0.027    &    12.254$\pm$0.013  &   a         &   0.866$\pm$0.001  &   0.19$\pm$0.02  &   0.056  &   -0.362$\pm$0.023   &   0.022$\pm$0.078   &    0.3$\pm$0.1  &    5.4$\pm$0.2  &  14  &  637.1   &       1.4$\pm$0.3  &    3.7$\pm$0.3  &   6.96E-05$\pm$1.31E-04   &  7.06E-11$\pm$1.52E-10  \\    
       V2  &   0.321549   &  64.498772  &  17.934$\pm$0.008    &    12.813$\pm$0.005  &   b         &   ---       ---    &   0.37$\pm$0.01  &   0.091  &    2.437$\pm$0.014   &  -0.180$\pm$0.038   &    0.1$\pm$0.1  &    0.7$\pm$0.1  &  11  &  596.5   &       3.0$\pm$0.1  &    9.6$\pm$0.1  &   9.23E-08$\pm$1.77E-02   &  2.24E-12$\pm$2.17E-19  \\    
       V3  &   0.241819   &  64.734032  &  13.230$\pm$0.009    &    12.450$\pm$0.073  &   a         &   ---       ---    &   0.25$\pm$0.10  &   0.078  &    0.153$\pm$0.089   &    ---       ---    &    0.6$\pm$0.1  &    5.6$\pm$0.3  &  11  &  102.9   &       1.1$\pm$0.1  &    4.2$\pm$0.1  &   1.39E-07$\pm$1.77E-02   &  1.21E-12$\pm$1.03E-03  \\    
       V4  &   0.426350   &  64.606903  &  15.262$\pm$0.014    &    14.495$\pm$0.043  &   a         &   0.971$\pm$0.001  &   0.50$\pm$0.06  &   0.131  &    0.241$\pm$0.062   &   0.036$\pm$0.053   &    6.4$\pm$2.2  &    3.1$\pm$0.2  &  10  &    0.4   &       4.8$\pm$2.3  &    3.4$\pm$0.3  &   8.53E-05$\pm$6.56E-04   &  1.84E-10$\pm$1.30E-09  \\    
 \end{longtable}
\endgroup
\vspace{ -0.7 cm}
\hspace{-8.5 cm}
a, b : Same as Table \ref{PMT} ;
c : Converted to $V$ and $I_{c}$ from SDSS data ;
$\dagger$ : Parameters derived from CMD ;
*  Parameters derived from SED \\

\end{scriptsize}

\begin{scriptsize}
\centering
\begingroup
\renewcommand\arraystretch{1}
\setlength{\tabcolsep}{2.2pt}
\begin{longtable}[c]{ccccccccccccccccccccc}
\caption{Sample of identified YSOs and their physical parameters as derived from their CMD and/or SED. The complete table is 
available in the electronic form only.\label{YSO}} \\
 \multicolumn{14}{}{Table}\\
 \hline
 
 ID  & $\alpha_{(2000)}$ & $\delta_{(2000)}$  &  $V$    &   $I_{c}$    &   Age$\dagger$   &  Mass$\dagger$         &  Flag   &  Data points*  &   $\chi^{2}$*  &    Age*     &  Mass*         &  Disc mass*            &  Disc accretion rate*  \\ 

  {}     & (degrees)   &  (degrees)   &  (mag)  & (mag)  & (Myr)       &  (M$_\odot$)        &       {}               &     {}        &            &    (Myr)  &  (M$_\odot$)  & (M$_\odot$) & (M$_\odot$~$yr^{-1}$) \\

 \endfirsthead
 
\hline
 \multicolumn{14}{c}{Continuation of Table \ref{YSO}}\\
 \hline 
 \hline
 \endhead
 
 \hline
 \endfoot
 
 \hline
 \multicolumn{14}{}{}\\
 
 \endlastfoot
\hline

Y1    &    0.386569   &  64.641273  &  16.840$\pm$0.020  &  14.282$\pm$0.014   &    0.1$\pm$0.1  &     1.3$\pm$0.1  &   a      &  13  &    8.9     &    0.3$\pm$0.2    &   4.5$\pm$0.8  &   4.23E-02$\pm$6.37E-02   &  4.57E-07$\pm$8.83E-07  \\   
Y2    &    0.399863   &  64.643867  &  19.324$\pm$0.064  &  17.113$\pm$0.022   &    1.4$\pm$0.3  &     1.0$\pm$0.1  &   a      &  13  &   34.5     &    0.1$\pm$0.1    &   4.0$\pm$1.3  &   4.56E-02$\pm$4.75E-02   &  5.57E-07$\pm$7.01E-07  \\   
Y3    &    0.405294   &  64.651543  &  18.795$\pm$0.013  &  16.674$\pm$0.008   &    1.1$\pm$0.2  &     1.2$\pm$0.1  &   a      &  13  &   20.0     &    3.8$\pm$2.4    &   3.5$\pm$1.2  &   5.63E-03$\pm$1.98E-02   &  7.84E-08$\pm$4.11E-07  \\   
Y4    &    0.676066   &  64.568512  &   ---       ---    &   ---       ---     &    ---     ---  &     ---     ---  & ---      &   5  &    0.1     &    1.4$\pm$2.0    &   2.0$\pm$1.3  &   7.92E-03$\pm$2.32E-02   &  8.25E-08$\pm$9.50E-07  \\   
 \end{longtable}
\endgroup
\vspace{ -0.75 cm}
\hspace{2.9 cm}
All the symbols are the same as in Table \ref{Variables} 
\end{scriptsize}
\end{landscape}
\clearpage
\twocolumn

\clearpage
\onecolumn
\begin{scriptsize}
\centering
\begingroup
\renewcommand\arraystretch{1}
\setlength{\tabcolsep}{8.0pt}
\begin{longtable}[c]{cccc}
 \caption{Classification and properties of 71 identified variables. 
The identification numbers are the same as Table \ref{Variables}.
 \label{variability nature}}\\
 \multicolumn{4}{}{Table}\\
 \hline
 
 {\bf ID} & {\bf Type} & {\bf Comment on Classification} & {\bf Comment on Variability}
 
 \endfirsthead
 
 \hline
 \multicolumn{4}{c}{Continuation of Table \ref{variability nature}}\\
 \hline  
 
 {\bf ID} & {\bf Type} & {\bf Comment on Classification} & {\bf Comment on Variability}
 \endhead
 
 \hline
 \endfoot
 
 \multicolumn{4}{}{}\\
 \endlastfoot
\hline

V1   &    PMS/Class\,{\sc iii}/WTT  &               CMD                       &  Periodic                                                                     \\
V2   &    PMS/Class\,{\sc iii}      &               CMD                       &  Non-periodic                                                                  \\
V3   &    PMS/Class\,{\sc iii}      &               CMD                       &  Non-periodic                                                                  \\
V4   &    PMS/Class\,{\sc iii}/WTT  &               CMD              &  Periodic                                                                      \\
V5   &    PMS/Class\,{\sc ii}       &             IR excess                  &  Periodic                                           
 \\
V6   &    PMS/Class\,{\sc ii}       &             IR excess         &  Periodic                                           
 \\
V7   &    PMS/Class\,{\sc iii}      &               CMD                       &  Non-periodic                                                                  \\
V8   &    MS/field                  &               CMD                       &  Periodic, new class of variable star                                                                      \\
V9   &    PMS/Class\,{\sc iii}/WTT  &               CMD                       &  Periodic                                                                      \\
V10  &    PMS/Class\,{\sc iii}/WTT  &               CMD                       &  Periodic                                                                      \\
V11  &    PMS/Class\,{\sc iii}      &               CMD                       &  Non-periodic                                                                  \\
V12  &    PMS/Class\,{\sc iii}      &               CMD                       &  Non-periodic                                                                  \\
V13  &    PMS/Class\,{\sc iii}      &               CMD                       &  Non-periodic                                                                  \\
V14  &    PMS/Class\,{\sc ii}/CTT   &           IR excess                    &  Non-periodic, few brightening/fading events visible in LC                                            \\
V15  &    PMS/Class\,{\sc ii}       &           IR excess                    &  Periodic                                                                      \\
V16  &    PMS/Class\,{\sc iii}/WTT  &               CMD                      &  Periodic                                                                      \\
V17  &    PMS/Class\,{\sc iii}/WTT  &               CMD             &  Periodic                                                                      \\
V18  &    PMS/Class\,{\sc iii}/WTT  &               CMD                       &  Periodic                                                                      \\
V19  &    PMS/Class\,{\sc ii}/CTT   &           IR excess                   &  Non-periodic, large scatter in LC $\sim$0.9 mag                                   \\
V20  &    PMS/Class\,{\sc ii}/CTT   &            IR excess             &  Non-periodic                                                                   \\
V21  &    PMS/Class\,{\sc iii}/WTT  &               CMD                       &  Periodic                                                                      \\
V22  &    PMS/Class\,{\sc iii}      &               CMD                  &  Non-periodic                                                                  \\
V23  &    PMS/Class\,{\sc iii}/WTT  &               CMD                  &  Periodic                                                                      \\
V24  &    PMS/Class\,{\sc iii}/WTT  &               CMD                 &  Periodic                                                                      \\
V25  &    PMS/Class\,{\sc iii}      &               CMD              &  Non-periodic                                                                  \\
V26  &    MS/field                  &               CMD                       &  Periodic                                                                      \\
V27  &    PMS/Class\,{\sc ii}/CTT   &           IR excess         &  Non-periodic, large variation in amplitude after 18$^{th}$ Oct 2017                                                                   \\
V28  &    PMS/Class\,{\sc iii}      &               CMD                      &  Non-periodic                                                                  \\
V29  &    MS/field                  &               CMD                       &  Periodic, $\delta$ Scuti star                                                   \\
V30  &    MS/field                  &               CMD                       &  Periodic                      \\
V31  &    PMS/Class\,{\sc ii}       &           IR excess                    &  Periodic, relatively longer period $\sim$9 days                                                                      \\
V32  &    PMS/Class\,{\sc iii}/WTT  &               CMD                       &  Periodic, large amplitude variation $\sim$1.4 mag                                                                      \\
V33  &    PMS/Class\,{\sc iii}/WTT  &               CMD                 &  Periodic                                                                      \\
V34  &    PMS/Class\,{\sc ii}       &           IR excess                    &  Periodic, both intra-night and inter-night variations present in LC      \\
V35  &    PMS/Class\,{\sc ii}/CTT   &           IR excess           &  Non-periodic, huge amplitude variation $\sim$2 mag  \\
V36  &    PMS/Class\,{\sc ii}/CTT   &           IR excess          &  Non-periodic, few brightening events visible in LC                                                                  \\
V37  &    PMS/Class\,{\sc iii}/WTT  &               CMD                &  Periodic                                                                      \\
V38  &    PMS/Class\,{\sc ii}       &           IR excess                    &  Periodic, relatively longer period $\sim$15 days 
\\
V39  &    PMS/Class\,{\sc iii}/WTT  &               CMD                       &  Periodic                                                                      \\
V40  &    MS/field                  &               CMD                       &  Very short period $\sim$4 hours                          \\
V41  &    ---                       &               ---                       &  Non-periodic                                                                  \\
V42  &    ---                       &               ---                       &  Non-periodic                                                                  \\
V43  &    MS/field                  &               CMD                       &  Non-periodic                                                                  \\
V44  &    MS/field                  &               CMD                       &  Periodic, large variation in amplitude  $\sim$1.6 mag.                       \\
V45  &    PMS/Class\,{\sc iii}/WTT  &               CMD                       &  Periodic                                                                      \\
V46  &    MS/field                  &               CMD                       &  Non-periodic, large variation in amplitude  $\sim$1.4 mag.                                                                  \\
V47  &    PMS/Class\,{\sc ii}/CTT   &           IR excess                    &  Non-periodic                                                                                                                                   \\
V48  &    MS/field                  &               CMD                       &  Periodic, $\beta$ Lyrae type LC                                                   \\
V49  &    PMS/Class\,{\sc iii}      &               CMD                       &  Non-periodic                                                                  \\
V50  &    MS/field                  &               CMD                       &  Periodic                                                                      \\
V51  &    MS/field                  &               CMD                       &  Very short period $\sim$4 hours                                                   \\
V52  &    PMS/Class\,{\sc iii}      &               CMD                       &  Non-periodic                                                                  \\
V53  &    PMS/Class\,{\sc ii}       &          IR excess          &  Periodic, relatively longer period $\sim$9 days                                                                                                                                           \\
V54  &    PMS/Class\,{\sc ii}       &          IR excess                     &  Periodic                                                                      \\
V55  &    MS/field                  &               CMD                       &  Periodic                                                                  \\
V56  &    MS/field                  &               CMD                       &   Very short period $\sim$4 hours                                                                            \\
V57  &    PMS/Class\,{\sc ii}       &          IR excess                     &  Periodic                                                                      \\
V58  &    PMS/Class\,{\sc iii}/WTT  &               CMD                       &  Periodic                                                                      \\
V59  &    MS/field                  &               CMD                       &  Periodic                                                                  \\
V60  &    PMS/Class\,{\sc iii}/WTT  &               CMD              &  Periodic                                                                      \\
V61  &    MS/field                  &               CMD                       &  Periodic, $\beta$ Lyrae type LC                                                                    \\
V62  &    MS/field                  &               CMD                       &  Periodic                                                                      \\
V63  &    PMS/Class\,{\sc iii}     &                CMD       &  Non-periodic                                                                  \\
V64  &    MS/field                  &               CMD                       &  Periodic, $\beta$ Lyrae type LC                                                                                                                                \\
V65  &    PMS/Class\,{\sc iii}/WTT  &               CMD                       &  Periodic, very short period $\sim$4 hours                               \\
V66  &    ---                       &               ---                       &  Periodic                                                                      \\
V67  &    PMS/Class\,{\sc iii}/WTT  &               CMD                       &  Periodic                                                                      \\
V68  &    ---                       &               ---                       &  Periodic                                                                      \\
V69  &    PMS/Class\,{\sc ii}       &           IR excess                  &  Periodic, long period $\sim$18 days, amplitude variation $\sim$1 mag                                                                                                                                           \\
V70  &    PMS/Class\,{\sc iii}      &               CMD                       & Non-periodic, amplitude variation $\sim$1 mag                                                                   \\
V71  &    ---                       &               ---                       &  Periodic, amplitude variation $\sim$1.1 mag                                                                      \\       

\hline
 \end{longtable}
\endgroup
\end{scriptsize}

\newpage
\clearpage
\onecolumn
\appendix
\section{Light curves of the variables not detected in $V$ band}
\label{appendix}

\begin{figure}
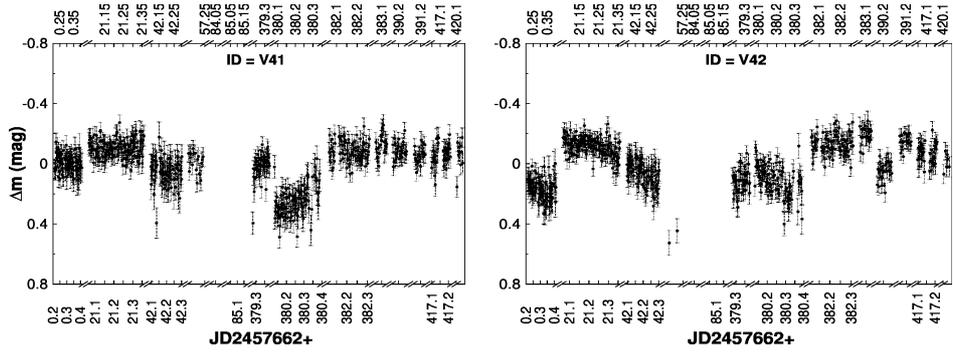

\centering
\includegraphics[width= 4.5 cm,height = 6 cm, angle=270]{figures/non-periodic/041-np.lreps} 
\hspace{0.2cm}
\includegraphics[width= 4.5 cm,height = 6 cm, angle=270]{figures/non-periodic/042-np.lreps}
\caption{LCs of two non-periodic variables.}
\label{Fig:additional_np}
\end{figure}

\begin{figure}
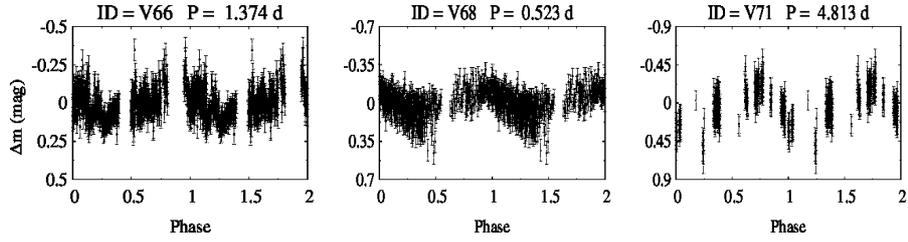

\centering
\includegraphics[width= 3.0 cm, height = 4.0 cm, angle= 270]{figures/periodic/066pha.lreps}
\hspace{0.2cm}
\includegraphics[width= 3.0 cm, height = 3.5 cm, angle= 270]{figures/periodic/068pha.lreps}
\hspace{0.2cm}
\includegraphics[width= 3.0 cm, height = 3.5 cm, angle= 270]{figures/periodic/071pha.lreps}
\caption{LCs of three periodic variables.}
\label{Fig:additional_periodic}
\end{figure}

\label{lastpage}
\end{document}